\documentclass[twocolumn]{aastex61}
\usepackage{dsfont}
\usepackage{amsmath}

\submitjournal{AJ}

\shorttitle{DARTTS-S I: SPHERE\,/\,IRDIS Polarimetric Imaging of 8 TTauri Disks}
\shortauthors{Avenhaus et al.}

\begin{document}

\title{Disks ARound TTauri Stars with SPHERE (DARTTS-S) I: SPHERE\,/\,IRDIS Polarimetric Imaging of 8 prominent TTauri Disks}

\correspondingauthor{Henning Avenhaus}
\email{havenhaus@gmail.com}

\author[0000-0002-1302-4613]{Henning Avenhaus}
\altaffiliation{Based on observations collected at the European Organisation for Astronomical Research in the Southern Hemisphere, Chile, under program 096.C-0523(A).}
\affiliation{Max Planck Institute for Astronomy, K\"onigstuhl 17, 69117 Heidelberg, Germany}
\affiliation{ETH Zurich, Institute for Particle Physics and Astrophysics, Wolfgang-Pauli-Strasse 27, CH-8093 Zurich, Switzerland}
\affiliation{Departamento de Astronom\'ia, Universidad de Chile, Casilla 36-D, Santiago, Chile}

\author[0000-0003-3829-7412]{Sascha P. Quanz}
\affiliation{ETH Zurich, Institute for Particle Physics and Astrophysics, Wolfgang-Pauli-Strasse 27, CH-8093 Zurich, Switzerland}
\affiliation{National Center of Competence in Research "PlanetS" (http://nccr-planets.ch)}

\author{Antonio Garufi}
\affiliation{Universidad Auton\'{o}noma de Madrid, Dpto. F\'{i}sica Te\'{o}rica, M\'{o}dulo 15, Facultad de Ciencias, 
E-28049 Madrid, Spain}

\author[0000-0003-2953-755X]{Sebastian Perez}
\affiliation{Departamento de Astronom\'ia, Universidad de Chile, Casilla 36-D, Santiago, Chile}
\affiliation{Millennium Nucleus ÒProtoplanetary DisksÓ Santiago, Chile}

\author{Simon Casassus}
\affiliation{Departamento de Astronom\'ia, Universidad de Chile, Casilla 36-D, Santiago, Chile}
\affiliation{Millennium Nucleus ÒProtoplanetary DisksÓ Santiago, Chile}

\author{Christophe Pinte}
\affiliation{Monash Centre for Astrophysics (MoCA) and School of Physics and Astronomy, Monash University, Clayton Vic 3800, Australia}
\affiliation{Univ. Grenoble Alpes, CNRS, IPAG, F-38000 Grenoble, France}

\author[0000-0001-5127-0172]{Gesa H.-M. Bertrang}
\affiliation{Departamento de Astronom\'ia, Universidad de Chile, Casilla 36-D, Santiago, Chile}

\author{Claudio Caceres}
\affiliation{Departamento de Ciencias Fisicas, Facultad de Ciencias exactas, Universidad Andres Bello, Av. Fernandez Concha 700, Santiago, Chile}
\affiliation{N\'ucleo Milenio Formaci\'on Planetaria - NPF, Universidad de Valpara\'iso, Av. Gran Breta\~na 1111, Valpara\'iso, Chile}

\author{Myriam Benisty}
\affiliation{Unidad Mixta Internacional Franco-Chilena de Astronom\'{i}a, CNRS/INSU UMI 3386 and Departamento de Astronom\'{i}a, Universidad de Chile, Casilla 36-D, Santiago, Chile.}
\affiliation{Univ. Grenoble Alpes, CNRS, IPAG, 38000 Grenoble, France}

\author{Carsten Dominik}
\affiliation{Astronomical Institute Anton Pannekoek, University of Amsterdam, The Netherlands}

\begin{abstract}

We present the first part of our Disks ARound TTauri Stars with SPHERE (DARTTS-S) survey: observations of eight TTauri stars {that were selected based on their strong (sub-)mm excesses} using SPHERE\,/\,IRDIS polarimetric differential imaging (PDI) in the $J$ and $H$ bands. All observations successfully detect the disks, which appear vastly different in size, {from $\approx$\,80~au in scattered light to $>$400~au}, and display total polarized disk fluxes {between 0.06\% and 0.89\%} of the stellar flux.
For five of these disks, we are able to determine the three-dimensional structure and the flaring of the disk surface, which appears to be relatively consistent across the different disks, {with flaring exponents $\alpha$ between $\approx$\,1.1 and $\approx$\,1.6}.
We also confirm literature results with regard to the inclination and position angle of several of our disks and are able to determine which side is the near side of the disk in most cases.
{While there is a clear trend of disk mass with stellar ages ($\approx$\,1~Myr to $>$\,10~Myr), no correlations of disk structures with age were found. }There are also no correlations with either stellar mass or sub-mm flux. We do not detect significant differences between the $J$ and $H$ bands. {However, we note that while a high fraction (7/8) of the disks in our sample show ring-shaped substructures, none of them display spirals, in contrast to the disks around more massive Herbig Ae/Be stars, where spiral features are common.}

\end{abstract}


\section{Introduction}

Significant progress has been made in recent years in providing empirical constraints on the physical and chemical properties of circumstellar disks, the cradles of future planetary systems, thanks to high spatial resolution observations. At (sub-)mm wavelengths, ALMA has been revolutionizing our understanding of the spatial distribution and properties of larger (millimeter-sized) dust grains, primarily found in the midplane of circumstellar disks, and of the molecular gas components \citep[e.g.][]{ALMA2015, andrews2016, huang2017, pinte2017, vanderplas2017}. At optical/near-infrared wavelengths, polarimetric differential imaging (PDI) observations with adaptive optics (AO) assisted, high-resolution and high-contrast cameras on 8\,m class telescopes have been yielding unprecedented images of the disks' surface layer by tracing scattering off the the smaller (micron-sized) dust grains \citep[e.g.][]{hashimoto2011, mayama2012, avenhaus2014b, avenhaus2014a, garufi2014, thalmann2015, akiyama2016, monnier2017, bertrang2018}, with the new SPHERE\,/\,IRDIS instrument being particularly successful at imaging disks at high signal-to-noise ratio 
\citep[e.g.][]{pohl2017, stolker2017}. An overview of recent SPHERE results can be found in \citet{garufi2017b}. Both techniques revealed a previously unknown richness and diversity in disk morphology and substructure. One of the key questions is to what extent these structures are leading to or are the result of planet formation processes. While the ALMA community has been publishing both papers investigating single sources in greater detail, as well as surveys with dozens of sources \citep[albeit with lower spatial resolution and sensitivity; e.g.][]{carpenter2014}, the high-contrast-imaging community was largely focusing on individual targets and, in addition, primarily on  Herbig Ae/Be stars \citep[e.g.][]{ohta2016, garufi2016, ginski2016, avenhaus2017}. There are ongoing activities starting to investigate larger samples of (Herbig Ae/Be) objects in order to understand evolutionary pathways \citep[e.g.][]{garufi2017a, ababakr2017}, but these studies are still rare. In addition, while some PDI studies also investigated the properties of TTauri disks \citep[e.g.][]{oh2016a, oh2016b, vanboekel2017}, disks around Herbig stars were easier targets, as they are generally larger in extent and brighter in scattered light. Furthermore, the generally brighter host star makes driving an AO easier. However, {while Herbig Ae/Be stars are more massive and hence more rare, TTauri stars (the progenitors of solar-like and lower-mass stars) are significantly more common}. In order to derive a comprehensive picture of circumstellar disk properties and identify correlations possibly related to disk evolution scenarios, larger samples across a wide range of stellar masses need to be studied in both scattered light and (sub-)mm emission.

Disks ARound TTauri Stars (DARTTS) is an effort at understanding TTauri disks, both in both scattered light and sub-mm, combining the power of SPHERE\,/\,IRDIS and ALMA to investigate disk structures at different wavelengths and similar high resolution.
This paper presents the results for the first eight sources of our DARTTS-S\footnote{Disks ARound TTauri Stars with SPHERE; PI: H. Avenhaus. The accompanying ALMA investigation of these disks under the DARTTS-A program is led by S. Perez} project, which is aimed presenting and analyzing a comprehensive NIR dataset of PDI observations of TTauri stars. It gives an overview of our results. {Part of the DARTTS-S data for DoAr~44 is presented and analyzed in detail in Casassus et al., (submitted), while further papers analyzing data for specific sources are in preparation.}

Thanks to its AO performance and sensitivity, VLT SPHERE\,/\,IRDIS is able to detect and reveal circumstellar disks even around low-mass stars with apparent magnitudes of R$\,\approx$\,10-13 mag. Here we focus on the first eight targets (see Table \ref{table:targetOverview}) and give a general overview of the observations, the data reduction (including a detailed description of the updated data reduction pipeline), and first quantitative results. Because the amount of data obtained is large, in-depth analysis and modeling of individual targets will be done in dedicated follow-up papers. However, the coherent observation technique and similar  signal-to-noise ratios (SNR) allow us to discuss first general trends and make comparisons across our target sample.

\begin{deluxetable*}{lccccccccc}
\centering
\tablecaption{Target overview
\label{table:targetOverview}}           

\tablehead{
\colhead{Target} & \colhead{alt. name} & \colhead{Sp. type} & \colhead{R [mag]} & \colhead{J [mag]} & \colhead{H [mag]} & \colhead{distance (pc)} & \colhead{f$_{\rm 1.3\,mm}$ [mJy]} & \colhead{$\rm \dot M$\,[M$_\odot$yr$^{-1}$]}
}
\startdata
IM~Lup		& Sz 82 & M0 & $\approx$\,10.8 & 8.783(21) & 8.089(40) & 158.45 \,$\pm$\, 1.34 & 200\,$^{[1]}$ & 1$\cdot$10$^{-11}$\,$^{\rm [I]}$\\ 
RXJ~1615	& RX J1615.3-3255 & K5 & 11.21 & 9.435(24) & 8.777(23) & 157.69 \,$\pm$\, 0.89 & 132\,$^{[2]}$ & 3$\cdot$10$^{-9}$\,$^{\rm [II]}$\\ 
RU~Lup		& Sz 83 & K7/M0 & $\approx$\,10.2 & 8.732(26) & 7.824(42) & 159.57 \,$\pm$\, 1.71 & 197\,$^{[3]}$ & 6$\cdot$10$^{-8}$\,$^{\rm [III]}$\\ 
MY~Lup		& PDS 77 & K0 & 11.06(5) &9.457(26) & 8.690(30)& 156.58 \,$\pm$\, 1.17 & 56\,$^{[4]}$ & $<$2$\cdot$10$^{-10}$\,$^{\rm [III]}$\\ 
PDS~66		& MP Mus & K1 & $\approx$\,10.0 & 8.277(32) & 7.641(23) & 98.86 \,$\pm$\, 0.30 & 224\,$^{[5]}$ & 1.3$\cdot$10$^{-10}$\,$^{\rm [IV]}$\\ 
V4046~Sgr	& Hen 3-1636 & K5+K7 & $\approx$\,10.3 & 8.071(23) & 7.435(51) & 72.41 \,$\pm$\, 0.34  & 283\,$^{[6]}$ & 2$\cdot$5$\cdot$10$^{-10}$\,$^{\rm [V]}$\\ 
DoAr~44		& V2062 Oph & K3 & 11.70 & 9.233(23) & 8.246(57) & 145.91 \,$\pm$\, 0.99 & 105\,$^{[7]}$ & 6$\cdot$10$^{-9}$\,$^{\rm [II]}$\\ 
AS~209		& V1121 Oph & K4 & $\approx$\,11.1 & 8.302(39) & 7.454(24) & 120.98 \,$\pm$\, 0.91 & 300\,$^{[8]}$ & 1.3$\cdot$10$^{-7}$\,$^{\rm [VI]}$\\ 
\enddata
\tablenotetext{}{Overview of our targets along with literature values. Spectral types and magnitudes are from SIMBAD. The $R$ magnitudes are given for reference, as the SPHERE AO is driven in the $R$ band. Where no $R$ magnitude is available, we roughly estimated it from the available magnitudes (indicated by "$\approx$"); however, all our targets are variable to some degree. {Note that V4046~Sgr is a spectroscopic binary and furthermore has a wide-separation binary companion \citep{kastner2011}}. Distances and 1$\sigma$ errors are from \citet{GAIA2018}. Additional distance references used in the writing of this paper: {\citet{krautter1997}, \citet{comeron2008}, \citet{torres2008}, \citet{andrews2011}. 1.3\,mm flux references: [1]~\citet{cleeves2016}, [2]~\citet{vandermarel2015}, [3]~\citet{vankempen2007}, [4]~\citet{lommen2010}, [5]~\citet{schutz2005}, [6]~\citet{rosenfeld2013}, [7]~\citet{nuernberger1998}, [8]~\citet{andre1994}. Accretion rate references: [I]~\citet{gunther2010}, [II]~\citet{manara2014}, [III]~\citet{alcala2017}, [IV]~\citet{ingleby2013}, [V]~\citet{donati2011}, [VI]~\citet{johnskrull2000}.}}

\end{deluxetable*}

\section{Our targets}

The first eight targets of our sample were selected based on (sub-)mm brightness. We chose to select those stars that have an extraordinarily high (sub-)mm flux, making sure to at the same time select stars covering a wide range of ages. The target list is thus not an unbiased selection of TTauri stars but a selection aimed at maximizing chances for detection. Some of the objects have previously been detected in scattered light. Literature values for the spectral types, distances and, $R$/$J$/$H$ band magnitudes, {as well as 1.3mm} photometry, can be found in Table \ref{table:targetOverview}. We derive age and stellar / disk mass estimates later in Section \ref{sec:ages} using pre-main-sequence tracks and sub-mm luminosities. Our targets in detail are as follows.

\subsection{IM~Lup}

IM~Lup is a well-studied M0 star located in the Lupus 2 cloud, classified as a weak-line TTauri star (WTTS) with weak accretion \citep{padgett2006, gunther2010}. 
It is a bright millimeter source as detected by SMA \citep{pinte2008} and ATCA \citep{lommen2007}, which indicates the presence of dust grains of several millimeters in size, with a dust mass of $\approx$\,10$^{-3}~$M$_\odot$.
The disk is inclined by 54$^\circ\pm3^\circ$ and can be traced in molecular gas emission to $\approx$\,750~au, with a break in the gas and dust density profile at $\approx$\,330~au \citep{panic2009}. 
Two rings are seen in the DCO$^+$ (3-2) line at radii of $\approx$\,320~au and $\approx$\,95~au, the inner of which can be connected to the CO snow-line, while the outer can be explained by non-thermal CO desorption at the position where the optical thickness of the disk decreases. Strong silicate features in the spectrum suggest the presence of micron-sized dust grains at the disk surface, which together with the millimeter data suggests spatial segregation of the dust grains as a function of size, for example from dust settling \citep{panic2009, oberg2011, oberg2015}.

The disk is revealed with Hubble Space Telescope (HST) scattered-light imaging, where the outer radius of the scattered-light disk can be shown to be $\approx$\,335~au, with a faint halo extending out to $\approx$\,700~au. Modeling of the system requires a flared disk (flaring exponent 1.13-1.17) with a scale height of 10~au at a distance of 100~au, and color measurements show a chromaticity of the disk between 0.6 and 1.6~$\mu$m which cannot be reproduced by simple scattering on spheres, suggesting the presence of aggregates on the disk surface \citep{pinte2008}. The latest available ALMA measurements show that the CO disk possibly extends even further, to $\approx$\,950~au in radius, making this one of the largest known protoplanetary disks with a disk mass of M$_{\rm gas}\,\approx\,0.17~$M$_\odot$ \citep{pinte2017}. These authors also confirm the sharp truncation of mm disk emission at smaller radii ($\approx$\,295~au), and show that it is also possible to directly measure radial and vertical temperature gradients in the disk.
All distances mentioned here have been scaled to the new \emph{Gaia} distance estimate of 158.45~pc (see Table \ref{table:targetOverview}). Several models for the available data exist in the literature \citep{pinte2008, panic2009, cleeves2016}.

\subsection{RXJ~1615}

RX~J1615.3-3255, which we abbreviate in this paper as RXJ~1615, is a WTTS located in an $\approx$\,1~Myr old part of the Lupus cloud
\citep{krautter1997, makarov2007}. It is identified as a transition disk, with modeling of high-resolution sub-mm data and \emph{Spitzer} IR spectroscopy pointing towards an inner hole extending clearly beyond the sublimation radius (based on the lack of near-IR excess) and a not fully cleared, but low-density cavity out to $\approx$\,25~au \citep{merin2010, andrews2011}. The latter authors also determine the total mass of the disk, assuming a gas-to-dust ratio of 100, to be as high as 0.128~M$_\odot$, {but the accretion rate ($4\cdot10^{-10}~$M$_{\odot}$yr$^{-1}$) is the lowest of all measured targets in their sample. Newer work gives a significantly higher accretion rate, though \citep[$3\cdot10^{-9}~$M$_{\odot}$yr$^{-1}$,][]{manara2014}.} The characteristic radius of the disk (as seen in the sub-mm continuum) is 98~au, with an inclination estimate of $\approx$\,41$^\circ$.

More recently, the disk was resolved through high-contrast imaging with VLT\,/\,SPHERE, both in polarization (with IRDIS and ZIMPOL PDI) and total intensity (using IRDIS and IFS ADI) by \citet{deboer2016}, who determined a disk inclination of $i=47^\circ\pm2^\circ$ and were able to resolve multiple rings at 1.50$\arcsec$, 1.06$\arcsec$, and 0.30$\arcsec$ ($237/167/47$~au), as well as another arc further out which they could not clearly determine to be either the rear surface of the disk or another ring. Earlier, \citet{kooistra2017} were able to image the disk using Subaru/HiCIAO PDI, albeit at significantly lower SNR, not being able to detect any of the disk rings and tracing the disk out to only $\approx$\,58~au. However, they were able to show that small dust grains must extend into the cavity seen in the sub-mm in order to be able to produce the scattered light signature seen in their observations and suggest that the small dust grain population must be radially decoupled from the larger grains. Neither of the observations was able to detect the inner gap in scattered light, despite the fact that the inner working angle in both cases was smaller than the $\approx$\,25~au of the sub-mm cavity size. All distances mentioned here have been scaled to the new Gaia distance estimate of 157.69~pc (see Table \ref{table:targetOverview}).

\subsection{RU~Lup}

RU~Lup is one of the most active and well-studied TTauri stars \citep{lamzin1996, stempels2002, herczeg2005}. This young object \citep[$\approx\,1$ Myr,][]{siwak2016} is located inside of the Lupus~2 cloud 
\citep{dezeeuw1999, comeron2008}. Its stellar mass is estimated to be slightly sub-solar \citep[$0.6-0.7~$M$_{\odot}$,][]{stempels2002} {with a high accretion rate of $6\cdot10^{-8}~$M$_{\odot}$yr$^{-1}$ \citep{alcala2017}}. Spectral line broadening as well as blueshifted emission line signatures indicate that RU~Lup is observed at a low inclination angle \citep{siwak2016}. 
The star exhibits both variations in radial velocity with a periodicity of $3.7\,$days, which was first interpreted as an indication for a $\approx$\,0.05~M$_{\odot}$ brown dwarf companion on a tight orbit by \citet{gahm2005}. However, the variations were later found to be more likely explained by the presence of large spots or groups of spots on the surface of RU~Lup itself, while a low-mass companion or stellar pulsations as source for these variations are discussed to be unlikely \citep{stempels2007}. Nevertheless, RU~Lup shows signs of an inner gap on au~scales which could be opened by a jupiter-like companion \citep{takami2003}. Its disk has not been imaged in scattered light before.

\subsection{MY~Lup}
MY~Lup is a K0 TTauri star located in the Lupus IV star forming region \citep{hughes1993, comeron2008, alcala2017}. 

It has been identified as a transition disk and a potential candidate for on-going planet formation \citep{romero2012}.

The disk has been observed previously by ALMA, where the inclination was determined to be $\sim$73$^\circ$ \citep{ansdell2016}, suggesting that it may partially be obscured by its circumstellar disk. Spectroscopic measurements have determined a remarkably low mass accretion rate as compared with similar disks in Lupus \citep{alcala2017, frasca2017}. This is consistent with the finding of a rather low gas-to-dust mass ratio from faint CO isotopologue ALMA observations \citep{miotello2017}. There are, so far, no studies of the disk in scattered light.

\subsection{PDS~66}

PDS~66 (also referred to as MP Mus) is a K1 classical TTauri star and one of the most nearby pre-main-sequence stars. 
The recent \emph{Gaia} measurement of $d=98.86\,\pm\,0.30$~pc \citep{GAIA2016} supports its membership in the $\epsilon$ Cha Association proposed by \citep{murphy2013}. 

The disk of PDS~66 was first imaged in scattered light with HST/NICMOS by \citet{cortes2009}, who estimated an outer radius of 170~au (with a distance estimate of 86~pc, translating to 195~au at the updated \emph{Gaia} distance) and an inclination of $32\pm5^\circ$. Their SED fitting suggested a disk inner edge at the dust sublimation temperature, though partial clearing may have happened already. More recent GPI (Gemini Planet Imager) images in PDI \citep{wolff2016} revealed a ring-like structure at 78~au separated from a bright inner disk by a 29~au wide region with diminished flux (radii have been updated with the new \emph{Gaia} distance).

The total dust mass of the disk is around 5$\cdot$10$^{-5}$M$_\odot$ \citep{carpenter2005}. A lower limit for the gas mass from CO measurements was given by \citet{kastner2010} at 9$\cdot$10$^{-6}$M$_\odot$, with the molecular gas disk extending out to $\approx$\,119~au (again converted using the new \emph{Gaia} distance estimate).

\subsection{V4046~Sgr}

V4046~Sgr is a close binary system, with two K-dwarfs of almost equal mass on a 2.4~day orbit \citep{quast2000, stempels2004}. {There is also a wide-separation (2.82$\arcmin$) binary that is likely loosely bound to the system \citep{kastner2011}}.
The SED in the IR shows a strong minimum between 5 and 8~$\mu$m, typical for transition disks,
and a silicate dust emission feature from large amorphous grains is present \citep[e.g.][]{rapson2015b}. 
Studies of the disk at 1.3~mm using ALMA reveal dust emission confined to a narrow ring centered at a radius of 37~au, 
with an central hole of a radius of $r=29$~au. This dust ring is embedded
in a larger CO gas disk with an inclination of $\approx$\,33.5$^\circ$, at a position angle of $\approx$\,76$^\circ$, and extending out to 300~au \citep{rosenfeld2012, rosenfeld2013}.
V4046~Sgr is a quite isolated young system at a distance of $\approx$\,73~pc. It is most likely a member of the $\beta$ Pic moving group \citep{torres2008},
and therefore about 23~Myr old \citep{mamajek2014}, making it the oldest system in our sample. V4046~Sgr is a special object: Not only is it the only gas-rich disk in the $\beta$ Pic moving group, but
also it resembles a Herbig Ae system in terms of the total mass of the two central objects, 
while in terms of luminosity it behaves like a TTauri system.

Disk images taken in polarized light by GPI were presented by \citet{rapson2015a}. These authors report a central cavity
inside $\approx$\,10~au, a ring with maximum flux around $\approx$\,14~au, and a gap at $\approx$\,20~au, 
as well as an outer halo extending to $\approx$\,45~au. The distances have not been re-calculated given the very small difference between the old distance estimate \citep[73 pc,][]{torres2008} and the new one \citep[72.41\,$\pm$\,0.34 pc,][]{GAIA2018}.

\subsection{DoAr~44}
DoAr~44 is a transition disk associated to $\rho$-Ophiuchus \citep{andrews2011}, and as such at a similar distance as AS~209, though \emph{Gaia} has not determined its distance individually.
Like most TTauri stars, it is actively accreting, and \citet{manara2014} have derived the accretion rate to be $\approx$\,6$\cdot$10$^{-9}$M$_\odot$yr$^{-1}$, one of the higher accrection rates amongst the 22 transition disks in their sample. 

The ALMA Band~7 continuum \citep[275-370~GHz / 0.8-1.1~mm,][]{vandermarel2016} reveals a fairly axially symmetric ring at a radius of 0.3$\arcsec$, which is inclined by $\approx$\,20$^\circ$ along a PA of $\approx$\,60$^\circ$. The total dust mass inferred from the continuum is 5$\cdot$10$^{-5}$M$_\odot$, while the gas mass inferred from the rare CO isotopologues is 2.5$\cdot$10$^{-3}$M$_\odot$.

{A subset of the DoAr~44 scattered light observations are presented in Casassus et al. (submitted), who propose a warped geometry to explain the polarized intensity.  Here, we place this object in context with the other sources.}

\subsection{AS~209}
AS~209 is a classical TTauri star \citep[spectral type K5,][]{perez2012} with a {high accretion rate of 1.3$\cdot$10$^{-7}\,$M$_\odot$yr$^{-1}$ \citep{johnskrull2000}}. The star is associated with the $\rho$-Ophiuchus cloud, but dwells in isolation from the main cloud members. AS~209 has a circumstellar disk which appears optically thin in continuum emission between 0.8-9.0~mm. The disk has a radius of $\approx$\,1" at 0.88~mm and becomes more compact at longer wavelengths. \citet{perez2012}  modeled these millimeter data finding evidence for radial variations of dust opacity at 0.2-0.5" resolution. It is inclined by $\approx$\,38$^\circ$ along a PA of $\approx$\,86$^\circ$ \citep{andrews2009}. 

ALMA observations of CO isotopologues report on a ring-like CO enhancement at $\approx$\,1$\arcsec$, possibly linked to CO desorption near the edge of AS~209's disk \citep{huang2016}. More recent data, also from ALMA \citep{fedele2017}, are able to identify two rings at $\approx$\,72~au and $\approx$\,124~au around a central core of emission, with gaps between them at $\approx$\,59~au and $\approx$\,98~au, at 2:1 resonance radii. The outer of these gaps is consistent with an approximately Saturn-mass planet opening it, while any planet in the inner gap would have to be less massive ($<$\,0.1~M$_{Jup}$). These ALMA data are also able to constrain the inclination and position angle more strictly, at 35.3$\pm$0.8$^\circ$ and 86.0$\pm$0.7$^\circ$, respectively.

There is no scattered light image of the disk available in the literature.

\section{Observations and data reduction}\label{observations_section}
\label{sec:observations}
\begin{deluxetable*}{lcccccccccc}
\centering
\tablewidth{0pt}
\tablecaption{Observation overview
\label{table:observations}}           
\tablehead{
\colhead{Target} & \colhead{Filter} & \colhead{DIT [s]} & \colhead{NDIT} & \colhead{NCYCLE} & \colhead{total frames} & \colhead{total time [s]} & \colhead{airmass} & \colhead{seeing [$\arcsec$]} & \colhead{$\tau_0$ [ms]} & \colhead{observation date}
}
\startdata
\vspace{-0.34cm}	& BB\_J & 64	& 2	& 7	& 56 (56)	& 3584 (3584) & 1.04-1.14 & 0.73-0.98 & 1.0-2.0 & March 11, 2016 \\
\vspace{-0.34cm}IM~Lup	& & & & & & & & \\
\vspace{ 0.14cm}	& BB\_H & 64	& 2	& 6	& 48 (48)	& 3072 (3072) & 1.07-1.16 & 1.07-1.52 & 2.7-4.1 & March 13, 2016 \\
\vspace{-0.34cm}	& BB\_J & 64	& 2	& 6	& 48 (48)	& 3072 (3072) & 1.16-1.34 & 0.88-1.26 & 1.3-3.0 & March 14, 2016 \\
\vspace{-0.34cm}RXJ~1615	& & & & & & & & \\
\vspace{ 0.14cm}	& BB\_H & 64	& 2	& 10.5	& 82 (80)	&5376 (5120) & 1.01-1.14 & 0.86-1.29 & 1.5-3.6 & March 14, 2016 \\
\vspace{-0.34cm}	& BB\_J & 64	& 2	& 9	& 72 (40)	& 4608 (2560) & 1.02-1.05 & 1.31-2.20 & 0.7-1.3 & March 11, 2016 \\
\vspace{-0.34cm}RU~Lup	& & & & & & & & \\
\vspace{ 0.14cm}	& BB\_H & 64	& 2	& 8	& 64 (64)	& 4096 (4096) & 1.04-1.13 & 1.08-1.47 & 1.6-2.7 & March 12, 2016 \\
\vspace{-0.34cm}	& BB\_J & 64	& 2	& 5	& 40 (40)	& 2560 (2560) & 1.05-1.08 & 0.79-1.07 & 1.9-2.9 & March 15, 2016 \\
\vspace{-0.34cm}MY~Lup	& & & & & & & & \\
\vspace{ 0.14cm}	& BB\_H & 64	& 2	& 5	& 40 (35)	& 2560 (2240) & 1.07-1.15 & 0.65-0.77 &  3.0-4.3 & March 15, 2016 \\
\vspace{-0.34cm}	& BB\_J & 64	& 2	& 6	& 48 (48)	& 3072 (3072) & 1.42-1.46 & 1.01-1.27 & 1.9-3.0 & March 14, 2016 \\
\vspace{-0.34cm}PDS~66	& & & & & & & & \\
\vspace{ 0.14cm}	& BB\_H & 64	& 2	& 7	& 56 (56)	& 3584 (3584) & 1.41-1.44 & 0.84-1.04 & 2.2-3.3 & March 15, 2016 \\
\vspace{-0.34cm}	& BB\_J & 64	& 2	& 6	& 48 (48)	& 3072 (3072) & 1.10-1.24 & 1.27-1.70 & 1.4-2.0 & March 12, 2016 \\
\vspace{-0.34cm}V4046~Sgr	& & & & & & & & \\
\vspace{ 0.14cm}	& BB\_H & 64	& 2	& 6	& 48 (48)	& 3072 (3072) & 1.05-1.15 & 0.89-1.19 & 1.8-2.7 & March 13, 2016 \\
\vspace{-0.34cm}	& BB\_J & 64	& 2	& 6	& 48 (48)	& 3072 (3072) & 1.01-1.08 & 0.72-0.86 & 2.8-4.5 & March 13, 2016 \\
\vspace{-0.34cm}DoAr~44	& & & & & & & & \\
\vspace{ 0.14cm}	& BB\_H & 64	& 2	& 5	& 40 (40)	& 2560 (2560) & 1.01-1.05 & 0.84-1.07 & 2.4-3.8 & March 15, 2016 \\
\vspace{-0.34cm}	& BB\_J & 64	& 2	& 8	& 64 (64)	& 4096 (4096) & 1.08-1.25 & 1.04-1.37 & 1.5-1.9 & March 10, 2016 \\
\vspace{-0.34cm}AS~209	& & & & & & & & \\
\vspace{0.14cm}	& BB\_H & 32	& 4	& 4	& 64 (64)	& 2048 (2048) & 1.01-1.02 & 0.73-0.91 & 2.7-3.6 & March 14, 2016 \\
\enddata
\vspace{.1cm}
\tablenotetext{}{Overview of our observations. The data were taken in PDI cycles, rotating through the four relevant half-wave-plate (HWP) positions. In each cycle, \emph{NDIT} integrations with an integration time of \emph{DIT} were taken before moving on to the next HWP position, for a total integration time of \emph{NDIT*DIT*}4 per cycle. A total of \emph{NCYCLE} of such cycles were taken, resulting in the total on-source integration time reported in the table. Since some frames were corrupted (for example because the adaptive optics could not stabilize the PSF), not all data were usable for all observations. The numbers in brackets represent the actual number of frames / integration time used in our data reduction. In case of the $H$-band observations of RXJ~1615, one cycle aborted after being half finished, resulting in a non-integer cycle number. Airmass, seeing and coherence time are as reported by the instrument. For the seeing, the IA detector linear fit estimate is reported.}
\end{deluxetable*}

All data presented in this paper were obtained during the six nights of March 10, 2016 to March 15, 2016, using SPHERE\,/\,IRDIS on the ESO VLT (Very Large Telescope). IRDIS was used in DPI mode in both $J$ and $H$ band, together with the N\_ALC\_YJH\_S coronagraph. Depending on the brightness of the source, either 32\,s or 64\,s integration times were used for the individual frames in order not to saturate the detector outside the coronagraph edge. Each observation followed the same pattern: flux frame (to measure the stellar flux and the PSF) - centering frame (to determine the exact position of the star behind the coronagraph) - science observations - second centering frame. No sky frames were taken. The total exposure times for the total of 16 observations (eight sources in two bands each) varied depending on the sky conditions and scheduling. The exact on-source times for each source/filter combination can be found in Table \ref{table:observations}.

The data reduction {(see appendix)} follows the general ideas presented in \citet{avenhaus2014a}, adapted for the IRDIS instrument and updated and improved where necessary. Specifically worth mentioning is the new way of correcting for instrumental polarization, which combines the equalizing of the ordinary and extraordinary beam with the technique of adding/subtracting scaled versions of the intensity frame to the Stokes Q and U parameters, pioneered by the SEEDS team \citep[e.g.,][]{follette2015}. Together with allowing for a polarized sky background component, this results in an overall better correction for instrumental effects.

There is self-cancellation flux loss close to the star due to the finite spatial resolution (i.e., the finite size of the PSF) combined with the fact that the local Stokes parameter Q$_\phi$ (or P) cannot be measured directly, but only the Stokes parameters Q and U can. This was first described in \citet{avenhaus2014b}, and is independent of the decomposition into the local Stokes vectors Q$_\phi$ and U$_\phi$, i.e. it occurs already in the Q and U frames. However, there are other side effects to this decomposition which have not been described in the literature yet. The patterns produced specifically in the U$_\phi$ image by these effects can closely resemble signals that one would expect from multiple scattering events \citep[e.g.,][]{canovas2015}, which means that it is easy to misinterpret them.

We describe both the origin of this effect, which we call Q$_\phi$/U$_\phi$ cross-talk, and the way we correct for it (at the same time correcting for self-cancellation close to the star), in the appendix, where we also describe the entire data reduction pipeline in detail again.

\section{Results and Analysis}\label{results}
\label{sec:results}

\begin{figure*}
\centering
\includegraphics[width=0.98\textwidth]{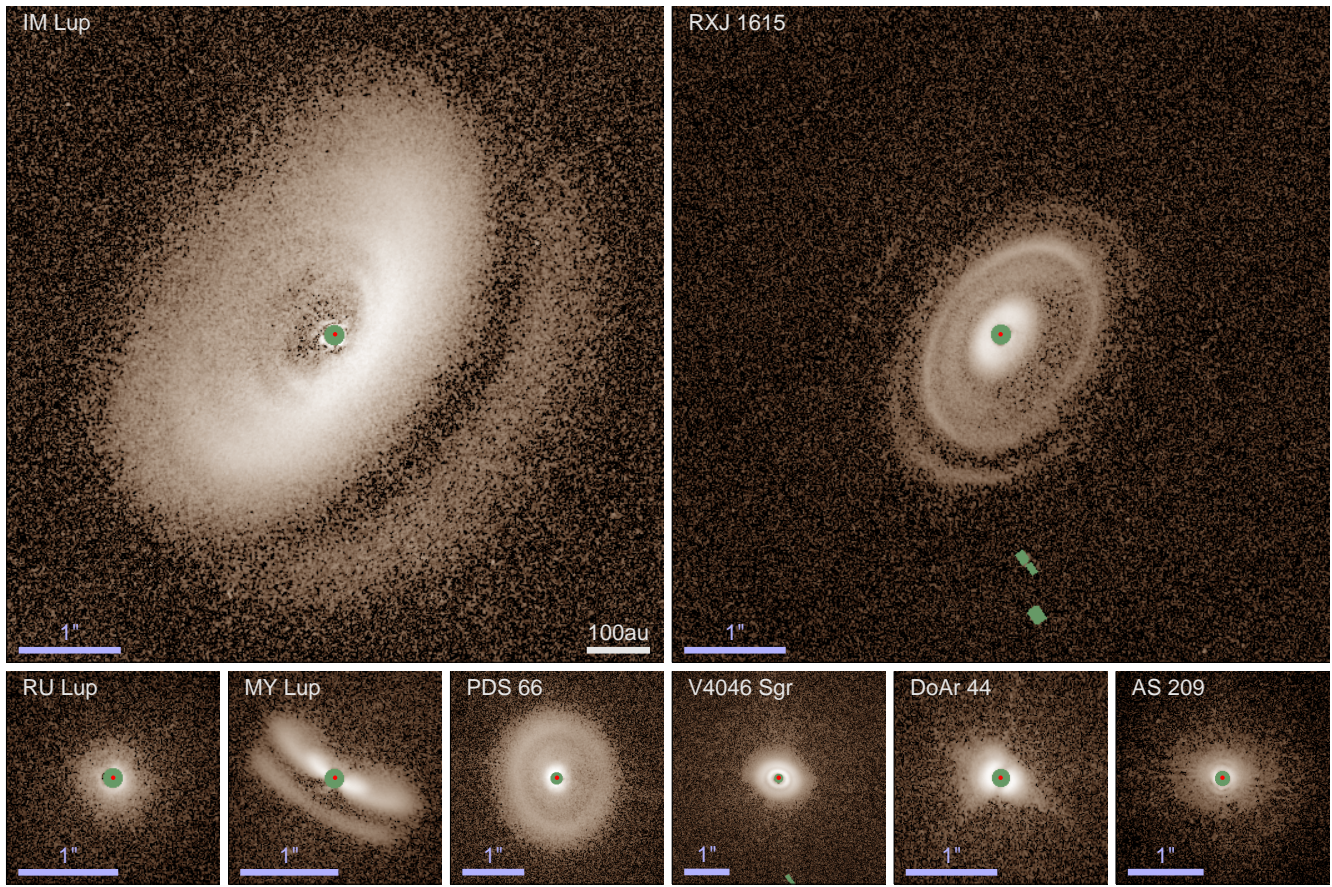}
\vspace{8pt}
\caption{H band images displayed in logarithmic stretch {(the exact stretch is adjusted for each disk individually to improve the visibility of substructures)}. The data were re-scaled to represent the same physical size, thus the 100\,au scale bar in the first panel applies for all panels. Because the angular scales are different, a 1\arcsec~bar is shown in each panel. Immediately obvious is the extraordinary size of the IM~Lup disk compared to the others, with RXJ~1615 coming in second. Areas marked green represent places where no information is available (due to either being obscured by the coronagraph or bad detector pixels). The red dot in the center marks the position of the star. North is up and east is to the left in all frames.}
\label{figOverview}
\end{figure*}

We successfully detect all eight TTauri disks in both $J$ and $H$ band, though the detection in $J$ band for some sources was only possible at low signal-to-noise ratio (SNR). We present {an overview of the higher-SNR $H$-band data for all eight disks, using logarithmic colormaps, in Figure~\ref{figOverview}. At the end of the paper, we also show our results in the more established way, where the data is multiplied by r$^2$  (Figure~\ref{figAll}). There, we also present the $J$ band and $U_\phi$ data}. The disks have been scaled in such a way that they represent the same physical scale. While this scale is afflicted with some uncertainty, due to the uncertainty in the distance specifically for the four sources with no \emph{Gaia} distance available, it is clear that the disks are of vastly different physical size, with IM~Lup being the largest and RU~Lup, almost identical in mass and of the same age, being one of the smallest. 

All disks except RU~Lup show easily visible substructure (see also Fig. \ref{figAll}). However, it is unlikely that the tightly spaced rings in AS~209 are real, because they only appear in the $H$ band and the depressions in the $Q_{\phi}$ image coincide with the diffraction rings in the intensity image. We discuss this in section \ref{secAS209Results}. There are, however, fainter structures in this disk that are hard to identify by eye, which we discuss in more detail in the same section.

\subsection{Surface brightnesses}

To get a first quantitative handle on the scattered light of the disks, we compare the brightness of their reflected, polarized light. Despite their different structures, inclinations, host star magnitudes and distances, we calibrate all our data with respect to the host star brightness. This way, we can compare how much of the incident starlight the disks reflect in total, keeping in mind that this figure is affected by the inclination of the disk. By comparing the $J$ and $H$ bands, we can get a rough estimate of the scattering color of the dust grains. Given the fact that we correct for the self-cancellation effect (as described above), we expect this figure not to be systematically affected by the difference in quality of the PSF between the $J$ and $H$ band. This figure also does not need to be corrected for distance, as both the stellar and the disk flux, as observed from Earth, scale the same with distance. We do have to keep in mind though that any parts of the disk that are behind the coronagraph, and their flux, cannot be accounted for.

\begin{deluxetable}{lccc}
\centering
\tablecaption{Ratio of polarized disk flux vs. stellar flux
\label{table:totalContrast}}           
\tablehead{
\colhead{Target} & \colhead{J band} & \colhead{H band} & \colhead{J/H ratio}
}
\startdata
IM~Lup	 	& 0.53\% $\pm$ 0.06\%	 & 0.66\% $\pm$ 0.05\%	 & 0.81 $\pm$ 0.12 \\
RXJ~1615	& 0.52\% $\pm$ 0.13\%	 & 0.67\% $\pm$ 0.32\%	 & 0.78 $\pm$ 0.42 \\
RU~Lup	 	& 0.06\% $\pm$ 0.03\%	 & 0.12\% $\pm$ 0.06\%	 & 0.51 $\pm$ 0.37 \\
MY~Lup	 	& 0.89\% $\pm$ 0.32\%	 & 0.81\% $\pm$ 0.27\%	 & 1.10 $\pm$ 0.55 \\
PDS~66	 	& 0.33\% $\pm$ 0.11\%	 & 0.26\% $\pm$ 0.06\%	 & 1.29 $\pm$ 0.52 \\
V4046~Sgr	& 0.46\% $\pm$ 0.18\%	 & 0.55\% $\pm$ 0.12\%	 & 0.85 $\pm$ 0.37 \\
DoAr~44	 	& 0.55\% $\pm$ 0.20\%	 & 0.65\% $\pm$ 0.24\%	 & 0.85 $\pm$ 0.45 \\
AS~209	 	& 0.18\% $\pm$ 0.07\%	 & 0.18\% $\pm$ 0.04\%	 & 1.02 $\pm$ 0.44 \\
\enddata
\end{deluxetable}

\begin{figure*}
\centering
\includegraphics[width=0.92\textwidth]{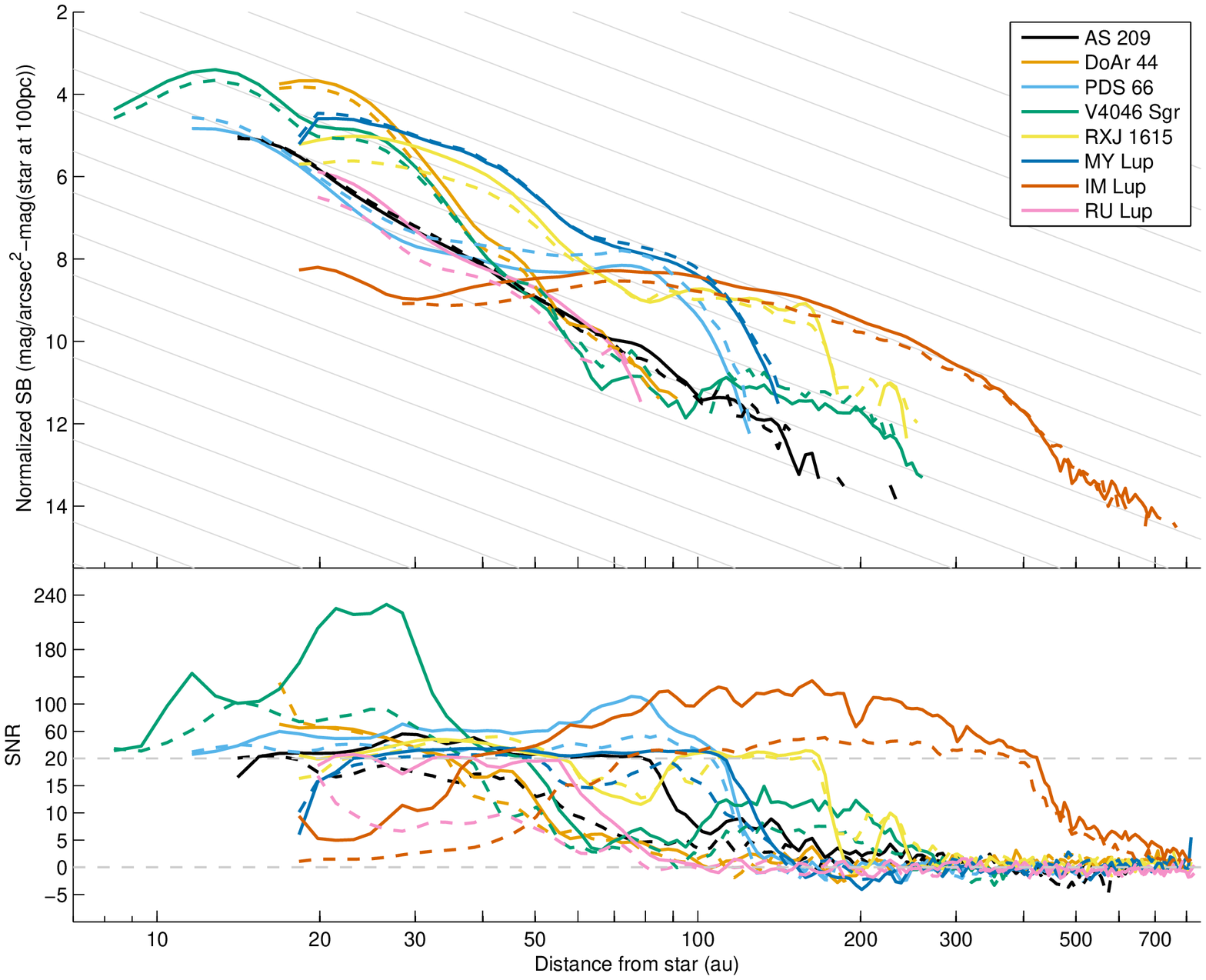}
\vspace{8pt}
\caption{Upper panel: Azimuthally averaged, normalized surface brightness versus distance from the host star for our targets, derived from the self-cancel-corrected images re-convolved with a 75\,mas Gaussian. Solid lines represent H~band, dashed lines J~band data. The width of the annuli used for averaging increases with radius proportional to r$^{1/2}$ (at 50~au, we use a width of 2.5~au).
For the sake of readability, error bars are omitted, and data are only shown where the detection is $>$\,3$\sigma$ or where the combined detection in $J$ and $H$ band is $>$\,3$\sigma$ and the detection in the individual band is $>$\,2$\sigma$. The lower panel shows the signal-to-noise ratio for all the data, with noise estimated from the U$_\phi$ frames. Note the change in scale at {SNR\,=\,20}. 
{Also note that even for our weakest detection, RU~Lup, the SNR peaks at $>$\,25$\sigma$. The significant negative SNR excursion at $\approx$\,500-600~au for AS~209 is to be discarded, it stems from time-variable striping of the IRDIS detector}. The gray background lines are for guiding the eye and scale as $r^{-2}$ (similar to the drop-off of stellar light with distance). {Note that errors or changes in the distance to the star, especially for those without GAIA measurements, would shift the curves along these background lines. Surface brightness plots in observational units, including surface brightnesses of the $U_\phi$ frames, can be found in Figure \ref{figSBAll}.}}
\label{figSB}
\end{figure*}

In Table \ref{table:totalContrast} we show the ratio between the reflected light of our disks and total intensity flux of the star-disk-system. We measure the polarized flux in an annulus between the edge of the coronagraph and a radius of 3.5$\arcsec$. Despite the fact that IM~Lup and RXJ~1615 and, to a lesser extent, PDS~66 appear significantly larger than the other disks, this does not mean that they reflect more light than, e.g., DoAr~44, one of the smallest disks in our sample. RU~Lup, also very small, is in fact also very faint, but the disk goes down to the coronagraph edge and more flux could be hidden from view under the coronagraph (the majority of the polarized flux usually comes from the innermost regions of the disk). The same is true for AS~209, which is also faint, but can actually be traced to about 200~au (see Figure \ref{figSB}). The third faintest disk, PDS~66, is also the third disk in our sample where it is known that the disk extends very close to the star.

The brightest disk in our sample, by this measure, is MY~Lup. However, this could be misleading as MY~Lup is highly inclined and the star likely shines partially through the disk, dimming the star (and thus decreasing the contrast between the star and the disk, making the disk relatively brighter). This interpretation goes well with the fact that the disk is apparently brighter in the $J$ than in the $H$ band - a reddening of the star due to dust extinction would have exactly this effect. It is also in line with the relatively high extinction of A$_{\rm v}$=1.2 (see Table \ref{table:derivedProperties}). However, our (conservative; see below) error estimates are large for these colors, such that essentially all disk colors agree with each other within the error bars. This evidence, just like the fact that all other disks except for PDS~66 are red, thus remains circumstantial.

It is important to keep in mind that the correction for self-cancellation we employ is a new technique, and depends on the quality of the PSF used. PSF fluctuations can thus cause over- or under-correction of the polarized flux, especially close to the coronagraph edge, potentially introducing errors. We are not able to estimate the quality of the PSF used for correction (which comes from the flux frames) compared to the mean PSF during the science observations in a meaningful way. We construct error bars by measuring the reflected light both in the uncorrected and corrected frames, and assume our errors to be smaller than the difference of the two measurements. This is a conservative error estimate, even though it does not take into account errors from e.g. the flux measurement of the star, as we expect those errors to be negligible compared to the effect the correction for self-cancellation has.

We also look at azimuthally averaged surface brightness curves (Figure \ref{figSB}). Again, we are aware that this does not take into account the inclination of the disk. In this case, we have to correct for the distance, because while the surface brightness is independent of distance, the stellar flux is not. We thus normalize the brightness of the disk (in mag/arcsec$^2$) with the magnitude of the star (as seen from 100~pc). Given the fact that we do only relative comparisons, we do not need to perform an absolute flux calibration of our data. The SNR for these surface brightnesses are determined from the variance in the U$_\phi$ images {(see appendix for a detailed description)}, and are shown separately in the bottom panel. We do not take errors that apply equally to all data points, such as errors in the flux measurement or distances to the stars, into account. For comparison, we also calculated the SNR from the variance in the Q$_\phi$ (rather than U$_\phi$) images. The maximum SNR determined in this way {is significantly lower ($\approx$\,35)} due to azimuthal flux variations in the Q$_\phi$ frames, but this effect is very much negligible in low-SNR regions, where there is not much flux to begin with. Thus, the regions where disk flux is detected at significant levels are virtually identical. Far out, the signal drops below the detection threshold for all disks, though this point is at $\approx$700-800~au for IM~Lup, which makes it by far the most extended disk in our sample. 

\subsection{Ring and spiral structures}

\label{sec:ringstructures}

Several of our disks show ring structures (best seen in Figure \ref{figAll}). In RXJ~1615, MY~Lup, PDS~66, and V4046~Sgr, full rings are seen, while DoAr~44 potentially shows a broken ring, resembling a smaller version of the HD142527 disk \citep{avenhaus2014a, avenhaus2017}, very close to the coronagraph edge (discussed in more detail in Casassus et al., submitted). IM~Lup shows several substructures in the $H$ band image which, at first sight, are hard to classify as either rings or a tightly wound spiral. In the $J$ band images, these substructures are washed out due to the lower Strehl. 

In order to investigate the rings in our data, we employ a method to automatically trace and fit the rings. In a first step, we de-project the data in order to be able to scale them by r$^2$, accounting for the drop-off of stellar illumination with distance. We then trace the ring at equally spaced position angles by fitting a 4th-order polynomial to the surface brightness in radial direction using a Markov Chain Monte-Carlo (MCMC) code in order to be able to determine the error in the position of the peak flux and re-project the fitted points into the image space. We use a second MCMC in order to fit the radius, inclination, position angle and h/r (i.e. the vertical offset off the midplane) of the ring. We assume the eccentricity of the rings to be zero. Because want to fit the ring in r$^2$-scaled surface brightness, we need to know the parameters of the ring in order to de-project for the first step of our routine. Thus, we start with an estimate for the parameters and iterate until convergence on a final solution is achieved. This allows us to also test the stability of our solution. The MCMC gives us access to statistical error bars for our parameters.

\begin{figure*}
\centering
\includegraphics[width=1\textwidth]{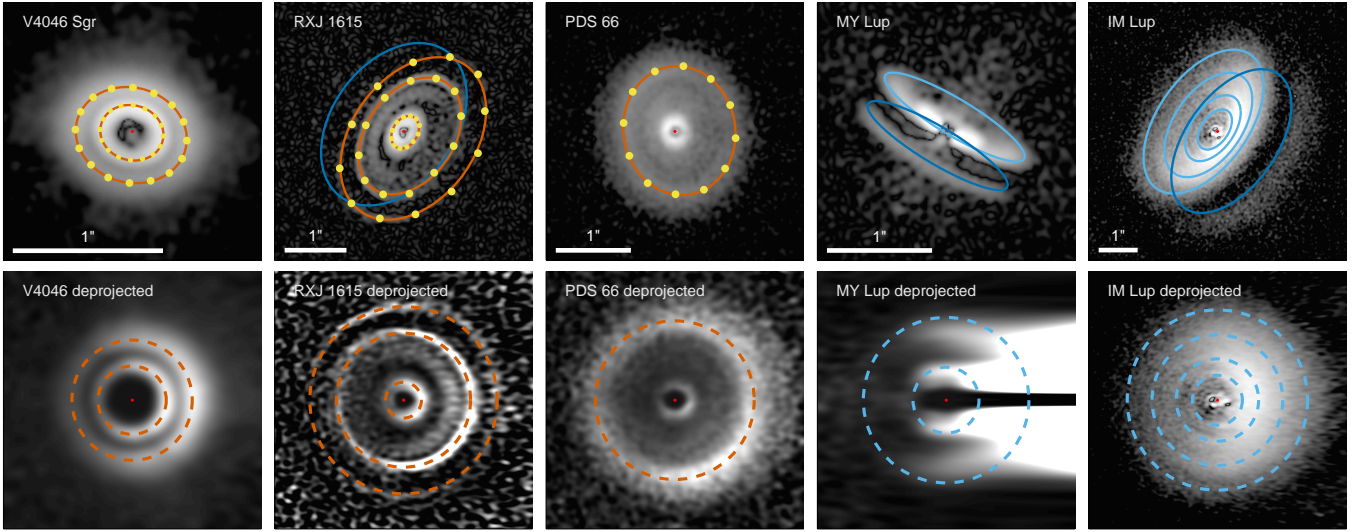}
\caption{Upper row: The disks of V4046~Sgr, RXJ~1615, and PDS~66 overlaid with their ring fits. For MY~Lup and IM~Lup, rings were overlaid by eye, because the automatic fitting procedure failed. Tracking points are yellow, ring fits are red and rings overlaid by eye are light blue. The rear edge of the disk (mirrored from the outermost ring) is shown in dark blue where applicable (MY~Lup, IM~Lup, RXJ~1615). Lower row: De-projected images of the disks, overlaid with their rings. We use flaring exponents of $\alpha$\,=\,1.605 (V4046~Sgr), $\alpha$\,=\,1.116 (RXJ~1615) and $\alpha$\,=\,1.271 (IM~Lup) for de-projection (see Section \ref{vertStructures}). For MY~Lup and PDS~66, where only one ring can be tracked, we use $\alpha$\,=\,1.2. In the de-projected image of MY~Lup, we additionally mark the approximate position of the second ring further in at 
r\,=\,0.31\arcsec / 46~au. For the de-projections, the semi-major axis is along the vertical, the semi-minor axis along the horizontal direction and the near side of the disk is always on the right. For the non-de-projected images, North is up and East is to the left.}
\label{figure:ringFits}
\end{figure*}

We do perform a number of checks to validate our results. First, we check whether it makes a difference whether we use a 4th- or 3rd-order polynomial to fit the position of the peak fluxes. Second, we visually check whether the fits of our rings coincide well with the location of the rings in the image (see Figure \ref{figure:ringFits}). Third, we start from a variety of initial guesses for the parameters and check whether we converge to the same solution (which is the case). We also check whether it makes a difference at how many azimuthal points we trace the ring (we tried using 8, 12, 16, 24, and 32 points, settling on 12 points for PDS~66 and RXJ~1615 and 16 points for V4046~Sgr, because convergence was reached fastest for these values).

For both V4046~Sgr and PDS~66, the ring fits are stable and agree within their error bars, independent of number of fitting points or polynomial order used. For these two disks we use the images corrected for self-cancellation of the disk (convolved with a 75\,mas FWHM Gaussian). The main difference in using these images compared to the uncorrected images is that the rings appear to be at smaller radii, specifically the rings close to the coronagraph edge (which makes sense given that the innermost regions are most affected by the self-cancellation). This effect is very minor, though ($<$\,5$\%$). The rings of RXJ~1615 are significantly more difficult to fit, and convergence is not reached for all numbers of tracking points. Also, using the corrected images makes the fits behave erratically and we thus choose to use the uncorrected data instead, where our method converges better. The problems mainly affect the h/r of the fit, which is unsurprising given the low SNR of the disk along the semi-minor axis. Asymmetries within the disk (see below) might also play a role.

We are able to fit three rings for RXJ~1615, two rings for V4046~Sgr, and the outer ring of PDS~66. RXJ~1615 clearly shows another ring between the first and second ring we track, but it is only seen on the northeastern side and we do not attempt to fit it. The broken ring of DoAr~44 is too close to the coronagraph for fitting to yield reliable results. The rings of MY~Lup are too inclined to allow for an automated tracking of the entire ring at all position angles, and the outer ring/edge of the disk of IM~Lup is so wide and diffuse that automatic tracking fails. We do, however, manually (by eye) overlay rings over these two disks, to get approximate estimates for their parameters.

For our fits, we assume the rings to be perfectly circular, i.e. they are not displaced from the center and thus have an eccentricity of zero. {We do not fit an offset of the ellipse from the stellar location, but the offset is intrinsically defined by the parameters we fit as:}

$$o_{\rm c} = R_{\rm ring}  \left(\frac{h}{r}\right) \sin{(i)}$$

\noindent {in the direction of (${\rm PA} + 90^\circ$). 
While it is possible that the rings do have an eccentricity or are not centered on the star, we do not find strong evidence for this. The rings are largely compatible with the errors of the fitted ring points, especially in the direction of the semi-major axis. }The possible exception are the rings of RXJ~1615 (discussed below), but our data is of too low SNR to reliably fit two additional parameters (eccentricity and position angle for the eccentricity), especially since these would be highly correlated with the inclination and h/r in our fits.

Our results are shown in Table \ref{table:rings} and Figure \ref{figure:ringFits}. For the rings where fitting is possible, the errors quoted are 1$\sigma$ errors from the MCMC fit. For the ones fit by eye, they are meant to represent approximate errors obtained by varying the parameters and seeing when they clearly do not fit any longer. There is no strong correlation between any of the variables except for the inclination and h/r, which are moderately correlated.

\begin{deluxetable*}{lccccc}
\centering
\tablewidth{16cm}
\tablecaption{Ring fits}
\tablehead{
\colhead{Ring} & \colhead{Radius (arcsec)} & \colhead{Radius (au)} & \colhead{Inclination} & \colhead{Pos. Angle} & \colhead{Flaring (h/r)}
}
\startdata
\vspace{-0.1cm}V4046~Sgr ring 1 & 0.212\,$\pm$\,0.001 & 15.35\,$\pm$\,0.06 & 30.53$^{\circ}\pm0.62^{\circ}$ & 74.40$^{\circ}\pm1.04^{\circ}$ & 0.093\,$\pm$\,0.006 \\
\vspace{0.1cm}V4046~Sgr ring 2 & 0.373\,$\pm$\,0.001 & 27.01\,$\pm$\,0.10 & 32.18$^{\circ}\pm0.51^{\circ}$ & 74.66$^{\circ}\pm0.72^{\circ}$ & 0.130\,$\pm$\,0.004 \\
\vspace{-0.1cm}RXJ~1615 ring 1 & 0.279\,$\pm$\,0.002 & 44.00\,$\pm$\,0.26 & 43.90$^{\circ}\pm1.12^{\circ}$ & 150.61$^{\circ}\pm0.94^{\circ}$ & 0.148\,$\pm$\,0.018 \\
\vspace{-0.1cm}RXJ~1615 ring 2 & 1.040\,$\pm$\,0.003 & 164.00\,$\pm$\,0.54 & 47.16$^{\circ}\pm0.87^{\circ}$ & 145.04$^{\circ}\pm0.48^{\circ}$ & 0.168\,$\pm$\,0.012 \\
\vspace{0.1cm}RXJ~1615 ring 3 & 1.455\,$\pm$\,0.013 & 229.44\,$\pm$\,1.99 & 46.78$^{\circ}\pm1.50^{\circ}$ & 143.82$^{\circ}\pm1.74^{\circ}$ & 0.183\,$\pm$\,0.020 \\
\vspace{0.1cm}PDS~66 ring 1 & 0.861\,$\pm$\,0.004 & 85.12\,$\pm$\,0.34 & 30.26$^{\circ}\pm0.88^{\circ}$ & 189.19$^{\circ}\pm1.33^{\circ}$ & 0.139\,$\pm$\,0.012 \\
\vspace{0.1cm}MY~Lup ring 1 ($^*$) & 0.77\,$\pm$\,0.03 & 120.57\,$\pm$\,4.70 & 77$^{\circ}\pm1.5^{\circ}$ & 239$^{\circ}\pm1.5^{\circ}$ & 0.21\,$\pm$\,0.03 \\
\vspace{-0.1cm}IM~Lup ring 1 ($^*$) & 0.58\,$\pm$\,0.02 & 91.90\,$\pm$\,3.17 & 53$^{\circ}\pm5^{\circ}$ & 325$^{\circ}\pm3^{\circ}$ & 0.18\,$\pm$\,0.03 \\
\vspace{-0.1cm}IM~Lup ring 2 ($^*$) & 0.96\,$\pm$\,0.03 & 152.11\,$\pm$\,4.75 & 55$^{\circ}\pm5^{\circ}$ & 325$^{\circ}\pm3^{\circ}$ & 0.18\,$\pm$\,0.04 \\
\vspace{-0.1cm}IM~Lup ring 3 ($^*$) & 1.52\,$\pm$\,0.03 & 240.84\,$\pm$\,4.75 & 55$^{\circ}\pm5^{\circ}$ & 325$^{\circ}\pm3^{\circ}$ & 0.23\,$\pm$\,0.04 \\
\vspace{0.1cm}IM~Lup ring 4 ($^*$) & 2.10\,$\pm$\,0.08 & 332.75\,$\pm$\,12.68 & 56$^{\circ}\pm2^{\circ}$ & 325$^{\circ}\pm2^{\circ}$ & 0.25\,$\pm$\,0.05 \\
\enddata
\tablenotetext{}{Results from fitting the rings present in our data. It is assumed that the rings are circular and displaced in vertical direction from the disk midplane. Note that this means thatThe $h/r$ parameter describes the height of the ring over the disk midplane, divided by the radius of the ring. This does not correspond directly to the gas scale-height of the disk, which we can not measure with our data, but the height of the last scattering surface. Note that the rings of IM~Lup and MY~Lup (marked with $^*$) are not fit using our procedure, but by eye. The radii in au are calculated using the distances to the stars and do not take into account the uncertainties in these distances, but only the statistical errors from the MCMC.}
\label{table:rings}
\end{deluxetable*}

Our ring fits are overlaid on the images in Figure \ref{figure:ringFits}, where we also show de-projected versions of our disks. For V4046~Sgr and PDS~66 there is no evidence for any asymmetries (such as breaks or deviations from circular structure) in the rings. RXJ~1615, on the other hand, shows some weak asymmetries, specifically in the second ring towards the southwest (upper left in the de-projected image), where the structure of the otherwise circular ring appears to be broken. This might be part of the reason for the fit being less stable than for the other disks. Furthermore, it is interesting that the theoretical rear edge of the disk (from mirroring the outermost ring to the back side) does not coincide with the faint ring arc that is seen towards the northeast of the disk. This might be because the actual disk is slightly larger than the outermost ring seen. A similar effect is seen in IM~Lup (towards the southwest).

Not surprisingly, the quality of the de-projections is lower for highly inclined disks, especially for MY~Lup, where the near side of the disk goes through the position of the star. However, de-projection still makes it possible to more clearly see the location of the second ring. What is also clear is that our naive visual fitting of the rings does not produce the correct radii of the ring, but rather fits the position of the outer edge.

\begin{figure}
\centering
\includegraphics[width=.48\textwidth]{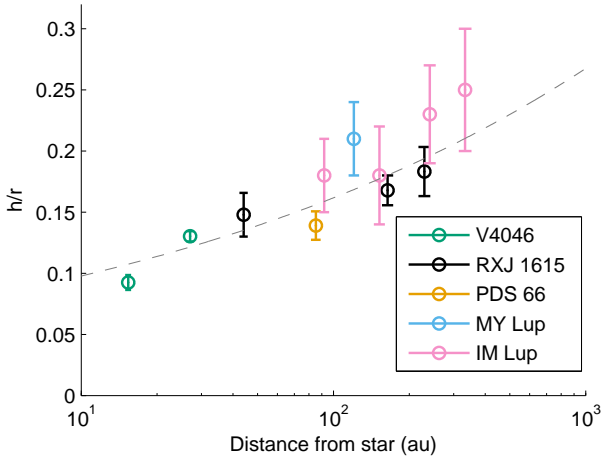}
\caption{Measurement of the h/r parameter of the rings we fit, plotted against the distance from the star. There is a clear trend towards higher h/r with larger distance. In fact, a single fit to all measurements {yields $\alpha$\,=\,1.219 and h/r\,(100\,au) = 0.1617} (grey dashed line). This fit is meant mostly to guide the eye, but works reasonably well for all disks except V4046~Sgr which, if considered separately, has a significantly higher flaring parameter.}
\label{figure:HoverR}
\end{figure}

\subsection{Vertical disk structures}

\label{vertStructures}

When looking at more than one ring, the behavior of h/r with radius can be described as a power law:

$$\frac{h}{r} = \frac{h_0}{r_0}\cdot\left(\frac{r}{r_0}\right)^{(\alpha-1)}$$

Where $h_0$ describes the $h/r$ value at a radius $r_0$, and $\alpha$ is the flaring index. $\alpha$ has to be higher than 1 in order to see the outer rings, because otherwise they would lie in the shadow of the inner rings. This also means that $h/r$ should be increasing with radius. We show all $h/r$ we measure in Figure \ref{figure:HoverR}, where it can be seen that $h/r$ clearly does increase with radius. 

Theoretical studies can derive flaring indices based on assumptions about the disk physics and geometry. For example, the \citet{chiang1997} model, by assuming a surface density profile $\sigma(r) \propto r^{-1.5}$, gives a temperature profile which translates into {a maximum flaring index of} $\alpha = \frac{9}{7} \approx 1.29$. For a thin disk model, with very small mass compared to the central star, the flaring is {expected to be $\alpha=\frac{9}{8}=1.125$ by \citet{kenyon1987}. The same authors derive the maximum flaring angle to be $\alpha=\frac{5}{4}=1.25$.}

In practice, we measure a flaring index of 1.605\,$\pm$\,0.132 in the case of V4046~Sgr, 1.116\,$\pm$\,0.095 in the case of RXJ~1615, and 1.271\,$\pm$\,0.197 for IM~Lup. These values (and errors) are acquired by fitting a power law to the h/r measurements of the rings and are only possible for disks where more than one ring can be measured. V4046~Sgr seems to be the clear outlier here, with a significantly higher flaring index, {inconsistent with the aforementioned theoretical values}. This is surprising given the fact that it is the oldest disk in our sample (and disks tend to settle with age). {The flaring of this disk could potentially be affected by the fact that V4046~Sgr is a K-dwarf spectroscopic equal-mass binary.} V4046~Sgr is also special in the sense that the rings we fit here are by far the closest to the star {and that it has a wide-separation binary companion \citep{kastner2011}}.

{If we use all data to fit the flaring behavior of our disks, we arrive at $\alpha=1.219\pm0.026$ and h/r\,(100~au)\,=\,0.1617\,$\pm$\,0.0051 (see Figure \ref{figure:HoverR}). While this is in reasonable agreement with theoretical studies \citep{kenyon1987, chiang1997}, it is difficult to interpret given the fact that all our systems are different and have no physical connection, and thus also no reason to show the same flaring, unless there is some intrinsic physical process that drives all disks towards a similar flaring behavior. We also have to remember that we can only measure the flaring of the last scattering surface using scattered-light data, and do not measure the flaring of the gas scale-height directly.}

\subsection{Disk rims and midplane shadows}

\begin{figure*}
\centering
\includegraphics[width=0.98\textwidth]{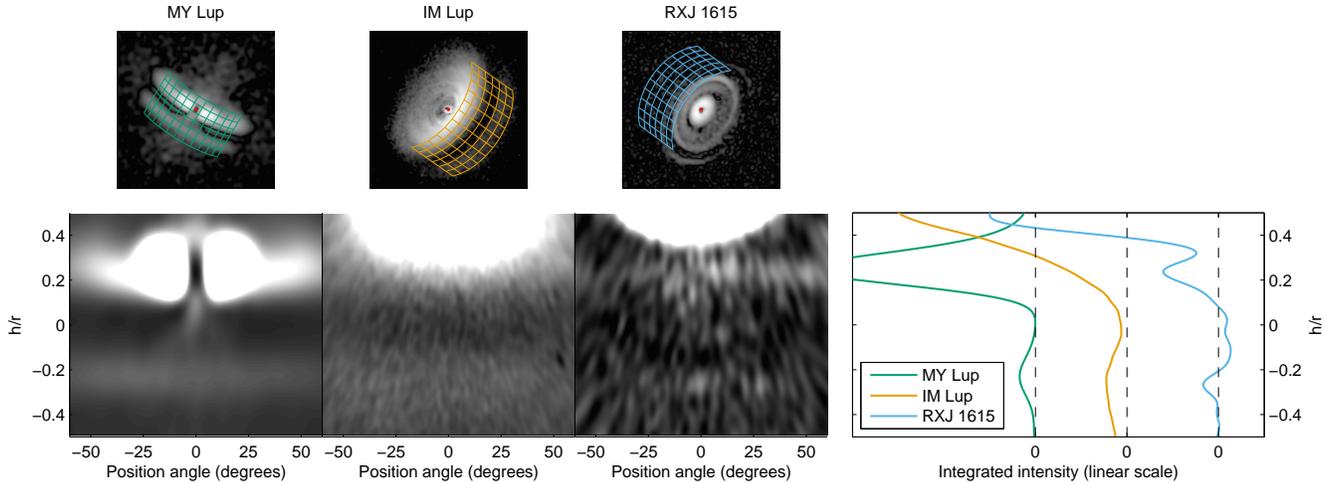}
\vspace{8pt}
\caption{{Left:} De-projection of the disk rims and surface brightness profiles perpendicular to the disk rim of MY~Lup, IM~Lup and RXJ~1615, {shown in linear scale}. The position angle refers to the position angle w.r.t. the disk minor axis. The meshes in the upper panels give a reference to show how the de-projection was done. {Right: Integrated intensities along the disk plane, between $-60^\circ$ and $+60^\circ$. The scaling of the data for the different disks with respect to each other is arbitrary. }As can be seen, the surface brightness goes into the negative for RXJ~1615, a sign that we over-corrected for self-cancellation.}
\label{figRimDeprojection}
\end{figure*}

For three of our disks (IM~Lup, RXJ~1615, and MY~Lup), the outer edge of the disk and thus the lower disk surface can be seen. To illustrate this, we de-project the $H$-band images of these outer rims for position angles from -60$^\circ$ to +60$^\circ$ around the disk minor axis. The resulting de-projections can be seen in Figure \ref{figRimDeprojection}. These de-projections use the data after correction for systematic self-cancellation and re-convolution with a Gaussian kernel (see Appendix). We use a 100\,mas FWHM kernel here in order to achieve slightly better smoothing for these faint features.

The de-projection shows that the two disk sides are parallel in all cases. However, it is not possible to estimate how dark the midplanes actually are. Without the discussed correction, the PSF convolution smears light from the disk upper and lower sides into the visible midplane gap. With correction, we can see that the midplane runs into negative values for RXJ~1615 (in fact, it partially does so for MY~Lup as well). This is a sign of an over-correction due to the application of a PSF that is worse than the average PSF encountered during the observations (this will be discussed in Avenhaus et al., in prep.). 

What can be seen, however, is that the two bright lanes on the disk rim of IM~Lup are relatively broad, significantly broader than the 100~mas kernel the data has been (re-)convolved with. For both MY~Lup and RXJ~1615, these features are much narrower (one has to remember that for MY~Lup, they are much closer to the star, thus the same h/r range is a smaller physical scale). This could mean that for IM~Lup, the disk at the location of the outer rim is optically thinner, allowing for deeper penetration of the stellar light and thus a larger range in heights above the midplane where light is scattered.

\section{Discussion}
\label{sec:discussion}

\subsection{Ages and and stellar / dust disk masses}

\label{sec:ages}

\begin{deluxetable*}{lcccccccc}
\centering
\tablecaption{Derived properties for our targets
\label{table:derivedProperties}}           

\tablehead{
\colhead{Target} & 
\colhead{T$_{\rm eff}$ [K]} &
\colhead{Av [mag]} & 
\colhead{Age [Myr]} &
\colhead{M$_\star$ [M$_\odot$]}& 
\colhead{L$_\star$ [L$_\odot$]} & 
\colhead{M$_{\rm dust}$ [M$_\Earth$]} & 
\colhead{f$_{\rm NIR}$/f$_\star$ [\%]} &
\colhead{f$_{\rm FIR}$/f$_\star$ [\%]} 
} 

\startdata
IM~Lup		&	4000		&	0.5	&	1.1$\pm$0.2	&	0.7$\pm$0.1				&	1.56\,$\pm$\,0.05					&		121$\pm$13	&	3\,$\pm$\,1		&		8\,$\pm$\,1	\\
RXJ~1615	&	4400		&	0.6	&	4.5$\pm$2.1	&	0.8$\pm$0.1				&	0.99\,$\pm$\,0.03					&		72$\pm$4		&	0\,$\pm$\,1		&		9\,$\pm$\,1	\\
RU~Lup		&	4000		&	0.0	&	1.2$\pm$0.4	&	0.7$\pm$0.2				&	1.44\,$\pm$\,0.13					&		95$\pm$6		&	40\,$\pm$\,4		&		30\,$\pm$\,2	\\
MY~Lup		&	5200		&	1.2	&	12.7$\pm$4.4	&	1.3$\pm$0.1				&	1.35\,$\pm$\,0.07					&		27$\pm$3		&	3\,$\pm$\,1		&		8\,$\pm$\,1	\\
PDS~66		&	5000		&	0.8	&	11.7$\pm$3.3	&	1.4$\pm$0.1				&	1.36\,$\pm$\,0.11 					&		43$\pm$3		&	7\,$\pm$\,1		&		7\,$\pm$\,1	\\
V4046~Sgr	&	4400		&	0.0	&	10.0$\pm$3.8	&	1.1$\pm$0.1\,/\,0.7$\pm$0.1	&	0.36\,$\pm$\,0.05\,/\,0.27\,$\pm$\,0.04	&		37$\pm$5		&	1\,$\pm$\,1		&		9\,$\pm$\,1	\\
DoAr~44		&	4800		&	2.2	&	6.8$\pm$2.4	&	1.5$\pm$0.1				&	1.88\,$\pm$\,0.17 					&		39$\pm$5		&	11\,$\pm$\,2		&		9\,$\pm$\,1	\\
AS~209		&	4600		&	0.8	&	1.9$\pm$0.9	&	1.1$\pm$0.1				&	1.75\,$\pm$\,0.07					&		78$\pm$8		&	7\,$\pm$\,2		&		19\,$\pm$\,2	\\
\enddata
\tablenotetext{}{Re-derived properties of our target stars and disks from our stellar modeling. {The effective stellar temperature is based on the spectral type of the star \citep[c.f.][]{cohen1979}, while} the extinctions are calculated from the SIMBAD colors and cross-checked with literature values. The relevant {spectral types and} distances can be found in Table \ref{table:targetOverview}. For targets without \emph{Gaia} distance, we assume an error of 20$\%$ in the distance estimate. The errors for the 1.3\,mm fluxes {(also found in Table \ref{table:targetOverview})} are typically small ($\sim$5$\%$) and do not dominate our error budget. For the visual extinction, we assume an error of 0.2\,mag. Descriptions of our derivations can be found in the main text.  }
\end{deluxetable*}

To place our dataset into context, we self-consistently calculate several stellar and disk properties. Half of our targets have new, accurate \emph{Gaia} distance estimates, which makes a re-calculation of the stellar ages particularly worthwhile, but to be consistent, we re-derive the properties for all sources in our sample. To do so, we retrieved the visible- to far-IR photometry for each source from SIMBAD and assumed a PHOENIX model of the stellar photosphere \citep{hauschildt1999} with solar metallicity, log($g$)~=~-4.0, and effective temperature $T_{\rm eff}$ obtained from the spectral of the star (found in Table~\ref{table:targetOverview}). We use the the relation described in \citet{cohen1979}. The choice of log($g$) is not critical, as within a range of reasonable values its impact on the stellar luminosity is marginal. Furthermore, we found self-consistency with the values of stellar mass and radius constrained at the end of this analysis. We de-reddened the observed photometry by means of the optical extinction $A_{\rm V}$ available from the literature and scaled the photospheric model to the de-reddened magnitude in the $J$ band. We then integrated the photospheric flux and converted it into the stellar luminosity $L_*$ using the distances found in Table~\ref{table:targetOverview}. Uncertainties on these estimates are primarily from $A_{\rm V}$, as well as the distance. We considered a $\Delta A_{\rm V}=0.2$ and the error from the \emph{Gaia} distance or a relative 20\% for sources without \emph{Gaia} data, and then propagated these uncertainties. Errors on $T_{\rm eff}$ and on the photometry are negligible in our error budget. Using the pre-main sequence tracks by \citet{siess2000}, we constrained the stellar age and mass as shown in Table~\ref{table:derivedProperties}. We take into account the fact that V4046~Sgr is a spectroscopic binary.

We also calculated the near- and far-IR excess of our sources similarly to \citet{garufi2017a}. These values were found by integrating the flux exceeding the stellar photosphere from 1\,$\mu$m to 5\,$\mu$m and from 20\,$\mu$m to 400\,$\mu$m, respectively. The relative uncertainties are given by the aforementioned uncertainty on $A_{\rm V}$.

Finally, we refined the estimate of the disk dust mass. To do so, we recovered the flux at 1.3 mm for all sources and scaled it as in \citet{beckwith1990} under the assumption that this emission is optically thin and by assuming a typical dust opacity of 2.3 cm$^2$g$^{-1}$ \citep[e.g.,][]{andrews2005} and a disk temperature of 25~$L_*/L_{\odot}$~K \citep[as in][]{andrews2013}, where $L_*$ is what we obtained above. The results are also shown in Table \ref{table:derivedProperties}.

\begin{figure}
\centering
\includegraphics[width=0.47\textwidth]{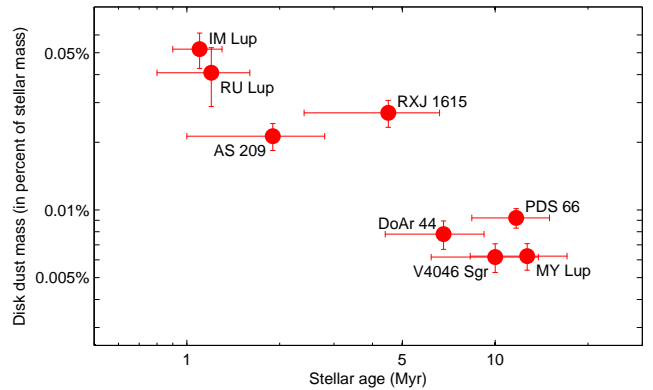}
\vspace{8pt}
\caption{Dust mass relative to stellar mass versus age of the source, {derived from literature data in order to put our data into context.} A trend towards lower fractional dust masses with higher ages is visible (as is to be expected). The two disks showing weak signals and no readily visible substructure (RU~Lup and AS~209) are interestingly among the youngest and most massive disks of our sample.}
\label{figAntonio}
\end{figure}

By comparing the obtained stellar ages and disk dust masses relative to their host star masses, we obtain the diagram found in Figure \ref{figAntonio}. While the error bars are large, the trend clearly points towards lower dust (disk) masses at advanced ages (as is to be expected). The three very young stars in our sample host a massive disk (in dust). This is somewhat surprising - while the fact that the disk of IM~Lup is young and massive can be expected just from looking at our scattered-light data, the other two very young sources (RU~Lup and AS~209) appear faint, compact, and feature-less in scattered light. At the same time, our calculations based on their 1.3mm fluxes shows that their disks must be massive. {While stellar age and dust mass seem correlated, there is no correlation between either parameter and disk substructure or total reflected light to be seen. However, it is worth pointing out that the two targets with faint disks (RU~Lup and AS~209) at the same time have the highest accretion rates amongst our sample (c.f. Table \ref{table:targetOverview}).}

Both also show very different SEDs from the rest of the sample. Their IR excess is in fact much more prominent (being 19\% and 30\% of the stellar flux) than the other objects ($\sim$\,7\%\,-\,9\%). In other words, a large amount of thermal reprocessing of the stellar light occurs around AS~209 and RU~Lup, and their dusty material is, along with IM~Lup and RXJ~1615, the most abundant among the sources of this work. RU~Lup also has the highest near-IR excess at 40\%, hinting at significant amounts of material close to the star.

The solution to this apparent incongruity is not obvious, but it is possible that the scarcity of detectable scattered light from the disk is related to a self-shadowing effect. This is the most likely explanation for the so-called Group II disks around Herbig Ae/Be stars, where the absence of a large disk cavity prevents the stellar light from reaching the outer disk regions \citep[see e.g.][]{garufi2017a}. Since we cannot probe the disk at separations of less than $\sim$100~mas, a lot of scattered light could be hidden in these innermost regions. In both disks, the disk extends down to the coronagraph, and no inner hole is detected, consistent with the literature, which shows that both disks extend close to the star \citep{takami2003, fedele2017}. However, the classifying criterion for Group II sources in the case of Herbig stars is the low far-IR excess \citep{meeus2001}, which is the opposite trend to AS~209 and RU~Lup. Furthermore, Herbig Group II stars are relatively old ($>$\,3 Myr) and their observations in PDI typically reveal either a compact but strong signal close to the star ($<$\,30 au) or nothing at all, whereas our data of AS~209 and RU~Lup show a relatively extended and faint signal.

We are thus more inclined to believe that the PDI data of AS~209 and RU~Lup reflect a geometry similar to RY Tau, which is another young TTauri star with a prominent IR excess but a relatively faint, diffuse, and feature-less signal in PDI \citep{takami2013}. According to these authors, this source would (still) be surrounded by optically thin and geometrically thick uplifted material which is entirely responsible for the observed scattered light and partly for the IR excess, whereas the underneath thin disk would only contribute to part of the IR excess. This explanation may hold for these two sources as well, and would also explain the absence (RU~Lup) or faintness (AS~209, c.f. section \ref{secAS209Results}) of disk features from our images in a framework where all PDI images of protoplanetary disks with sufficient signal-to-noise ratio available from the literature show some sort of substructures - except, to our knowledge, RY~Tau.

\subsection{Individual targets}

\subsubsection{IM~Lup}

Our results for IM~Lup are most readily compared to those derived by \citet{pinte2008}. These authors imaged the disk of IM~Lup with the Hubble Space Telescope in the visible and near-IR. Their results in terms of disk position angle are consistent with our results: $143\pm5^\circ$ (compared to our estimate of $325/145\pm2^\circ$). They are also able to detect the lower surface of the disk as well as the dark lane between the two disk surfaces, results that we can confirm at much higher signal-to-noise ratio. Our results additionally allow us to detect substructure in the disk upper surface, although it is not entirely clear whether this substructure represents rings or tightly wound spiral structures. We are also able to trace the disk to much smaller angular separations. \citet{pinte2008} describe a faint halo out to $\approx$\,4.4$\arcsec$, which they ascribe to a tentative optically thin envelope around the disk. In our surface brightness profile (Fig.~\ref{figSB}), it can be seen that while the azimuthally averaged surface brightness drops steeply beyond 400-450~au (2.5-2.8$\arcsec$), the signal can be detected out to $>$\,700~au (4.3$\arcsec$). The signal in this region is very faint and cannot be seen directly in the images, but only when azimuthally averaging. It is also relatively close to the edge of the detector, where various imperfections occur. However, none of the other sources show consistent signal in both $J$ and $H$ band at these angular separations, thus we conclude that this signal is indeed real. However, the faint envelope must be optically thin given that the back side of the disk can be seen through it.

The disk is modeled with a gas pressure scale height of 10~au at a radius of 100~au with a flaring index of $1.13-1.17$ by \citet{pinte2008} ($h/r=0.1$). This can be compared to our estimate of $h/r=0.18\pm0.03$ at a similar radius (see table \ref{table:rings}), which suggests that the $\tau=1$ scattering surface resides at around 1.8 pressure scale heights. Our estimate for the flaring index is much less well-constrained at $1.27\pm0.20$ (also remember that the fitting was done by eye), but consistent with these results.

\citet{panic2009} describe Submillimeter Array (SMA) data of the source, with which they are able to determine Keplerian rotation of the disk in clockwise direction (as seen from our vantage point). They furthermore constrain the disk inclination to $54\pm3^\circ$, consistent with our estimate of $56\pm2^\circ$ for the outermost ring. The disk can be traced in the gas out to 900~au (assuming a distance of 190~pc, translating to 751~au at the \emph{Gaia} distance of 158.45~pc, similar in radius to our scattered-light observations), while the continuum observations can trace the dust only to around 400~au (334~au given the updated distance). Their model thus requires a break in the disk surface density at around this distance, which is consistent with the scattered light observations, which show that the disk is truncated relatively sharply beyond 400~au (2.5$\arcsec$), with the outermost ring we trace at $2.1\pm0.08\arcsec$ ($333\pm13$~au). 

More recently, two tentative dust rings have been detected at millimeter wavelengths by ALMA \citep{cleeves2016, pinte2017}, with radii of $\approx$\,150\,au and $\approx$\,250\,au, i.e. approximately where we observe the faint rings \#2 and \#3 in our $H$-band image. The current resolution of the ALMA observations of IM Lupi ($\approx$\,0.3$\arcsec$) is insufficient to determine if there is any structure at millimeter wavelength that could be associated with our ring \#1.The millimeter emission drops sharply outside of 310\,au and no emission is detected at the location of our ring \#4.

\citet{pinte2017} use the individual channel maps of the CO isotologues to determine the altitude of the emitting layers. Interestingly, the scattered light $\tau=1$ surface we measure ($h/r \approx 0.2$ around 200\,au) appears to be located between the $^{13}$CO ($h/r \approx 0.16$) and the $^{12}$CO ($h/r \approx 0.35$) at the same radii. No evidence of structure has been detected in the CO map, possibly due to the limited spatial resolution of the observations ($\approx$\,0.3$\arcsec$). Observations of the CO emission at higher spatial resolution could potentially detect the counterpart of the structures we see in the SPHERE data and shed some light on their nature, in particular on their kinematics.

\subsubsection{RXJ~1615}

The disk of RXJ~1615 was previously detected and described by \citet{deboer2016}, using both SPHERE\,/\,IRDIS H2H3 dual-band ADI (angular differential imaging) as well as IRDIS $J$-band and ZIMPOL R-band PDI. They clearly detect rings 2 and 3 we describe (which they call R2 and R1, respectively), along with an arc inside of those two rings (A2), which we describe here as well. This arc is most likely a full ring, which we are able to trace for more than $180^\circ$ (see Figure~\ref{figRimDeprojection} and discussion in section \ref{sec:ringstructures}). They describe an elliptical inner disk component, which we fit here as our ring 1. 

They also describe an arc outside the outermost ring and discuss whether it is another ring, or the back side of the disk. Given our higher-SNR polarimetric observations, we are convinced that this is indeed the disk back surface, even though it is not at the location of the projected outermost ring of the front surface of the disk. This can be explained by the fact that the light has to a) 'bend around' the disk edge to reach us from the disk back surface, and that b) the truncation radius of the disk must not necessarily coincide with the radius of the outermost surface ring (ring 3 in our discussion). Besides the fact that the geometrical structure in both our images and de-projections (see Figure~\ref{figRimDeprojection}) as well as the data shown in \citet{deboer2016} seem more consistent with this geometry, we would expect a 4th ring to be most easily detected along the disk major axis (where the SNR for all other rings is highest) rather than the disk minor axis. We thus strongly favor the back surface explanation over the additional ring.

\begin{deluxetable}{llccc}
\centering
\tablecaption{Comparison of ring fits for RXJ~1615
\label{table:RXJ1615}}           

\tablehead{
\colhead{\#} & \colhead{Par.} & \colhead{this work} & \colhead{ADI-H23} &\colhead{PDI-J} 
}
\startdata
\vspace{-0.1cm}1			&	R [$\arcsec$] 		&	0.279\,$\pm$\,0.002 	&		0.30\,$\pm$\,0.01 	&	0.35\,$\pm$\,0.01	\\
\vspace{-0.1cm}	 		&	incl. [$^\circ$] 		&	43.9\,$\pm$\,1.1 	&		49.0\,$\pm$\,3.9 	&	47.7\,$\pm$\,4.1	\\
\vspace{-0.1cm}	 		&	PA [$^\circ$] 		&	150.6\,$\pm$\,0.9 	&		145.4\,$\pm$\,4.2 	&	144.5\,$\pm$\,4.3	\\
\vspace{0.1cm}	 		&	h/r				&	0.148\,$\pm$\,0.018 	&		n/a			 	&	n/a				\\
\vspace{-0.1cm}2			&	R [$\arcsec$] 		&	1.040\,$\pm$\,0.003 	&		1.06\,$\pm$\,0.01 	&	1.06\,$\pm$\,0.01	\\
\vspace{-0.1cm}	 		&	incl. [$^\circ$] 		&	47.2\,$\pm$\,0.9 	&		48.5\,$\pm$\,1.3 	&	46.8\,$\pm$\,1.4	\\
\vspace{-0.1cm}	 		&	PA [$^\circ$] 		&	145.0\,$\pm$\,0.5 	&		145.4\,$\pm$\,1.3 	&	144.3\,$\pm$\,1.4	\\
\vspace{0.1cm}	 		&	h/r		 		&	0.168\,$\pm$\,0.012 	&		0.158\,$\pm$\,0.014 	&	0.152\,$\pm$\,0.013	\\
\vspace{-0.1cm}3			&	R [$\arcsec$] 		&	1.455\,$\pm$\,0.013 	&		1.50\,$\pm$\,0.01 	&	1.50\,$\pm$\,0.01	\\
\vspace{-0.1cm}	 		&	incl. [$^\circ$] 		&	46.8\,$\pm$\,1.5 	&		47.3\,$\pm$\,1.0 	&	47.0\,$\pm$\,0.8	\\
\vspace{-0.1cm}	 		&	PA [$^\circ$] 		&	143.8\,$\pm$\,1.7 	&		145.7\,$\pm$\,1.0 	&	144.2\,$\pm$\,0.8	\\
\vspace{0.1cm}	 		&	h/r			 	&	0.183\,$\pm$\,0.020 	&		0.162\,$\pm$\,0.009 	&	0.162\,$\pm$\,0.007	\\
\enddata
\tablenotetext{}{Comparison between the ring parameters derived in this work and those derived by \citet{deboer2016} using $H$-band ADI (column ADI-H23) and $J$-band PDI (column PDI-J).}
\end{deluxetable}

\citet{deboer2016} also fit ellipses to the rings and the inner disk. We compare their to our fits in table \ref{table:RXJ1615}. Even though these results do not take into account systematic errors, they agree within the error bars in terms of inclination, position angle and flaring (h/r for the $\tau=1$ surface). In terms of radii, the results do not agree, but as we pointed out in section \ref{sec:ringstructures}, the determination of the radii is slightly arbitrary and depends on the exact definition of where you place the peak of the ring.

Our results are also reasonably close to the results from \citet{vandermarel2015}, who obtain $i=45^\circ$ and $PA=153^\circ$, though do not state errors for these measurements. They model the disk with a cavity radius of 17~au, a characteristic radius (for the exponential taper of the continuum) of 98~au, and an outer radius of 170~au (distances updated using the new Gaia measurement). From our results, we can see that the cavity in small dust grains must be smaller, as we detect scattering down to the coronagraph edge ($\approx$\,0.1$\arcsec$ / 15.8~au). We also see that in scattered light, the disk is much larger than 170~au, as the outermost ring is detected at $\approx$\,230~au, with the outer edge of the disk likely a bit further out.

\subsubsection{RU~Lup}

RU~Lup is the most unremarkable disk in our sample. While the star shares many characteristics with IM~Lup (in terms of age, spectral type, sub-mm excess, SED), the two disks appear completely different in scattered light. RU Lup is both the faintest and reddest disk in our sample (c.f. table \ref{table:totalContrast}), though the second measurement could be impacted by bad observing conditions and the fact that both the star and the disk are faint. The disk appears to be brighter in both the $J$ and $H$ band in the south-west direction. This could be a hint towards a low to moderate inclination along the SE-NW-axis with the SW side being the near side, {but this interpretation is speculative (see also the discussion on AS~209 in Section \ref{secAS209Results}).}

RU~Lup is known to have a rather high accretion rate of $\sim10^{-7}$M$_\odot$yr$^{-1}$ \citep{podio2007}. This could be related to the disk extending very far in and not showing any signs of an inner gap in our scattered light observations. Archival SPHERE/ZIMPOL data show the disk to extend in to at least $\sim$40~mas, and \citet{anthonioz2015} resolve the disk using VLTI/PIONIER and fit it with an inner radius of $\sim$0.1~au (0.7~mas). ATCA measurements indicate a Gaussian size of the disk of 1.02$\pm$0.32$\arcsec$ at 1.4~mm \citep{lommen2007}, somewhat smaller than the same measurement for IM Lup (1.33$\pm$0.20~$\arcsec$).

The most likely explanation for our observations is that the disk of RU~Lup is extending very close to the star and is not very flared, such that partial shadowing reduces the amount of light reaching the outer parts of the disk, and thus the amount of scattered light to be detected. Whether substructures are present in the disk can not be determined due to the low signal-to-noise ratio.

\subsubsection{MY~Lup}

MY Lup is the most highly inclined disk in our sample, but otherwise resembles the structure of the IM~Lup and RXJ~1615 disks at smaller size (flared, truncated, with multiple rings on the surface) even though the disk is significantly older than those two targets. Its age has previously been determined to be 16 Myr by \citet{frasca2017}, which is consistent with our estimate of 12.7\,$\pm$\,4.4 Myr, however its spectrum seems to be under-luminous compared with young TTauris with similar spectral types, thus a younger age is still quite plausible.

This issue is likely related to the high inclination. Because of the disk being so inclined, some of the starlight is obscured by the circumstellar disk. This was previously pointed out by \citet{ansdell2016} based on their estimate of an inclination of $\sim$73$^\circ$, and is confirmed by our observations which clearly show the disk being highly inclined at $\sim$77\,$\pm$\,1.5$^\circ$ and indicate obscuration by the outer ring. This also ties in to the disk appearing blue w.r.t. the starlight, as the starlight is most likely reddened because of being filtered by the dust disk.

\subsubsection{PDS~66}

Our images confirm the overall morphology of the GPI and HST images \citep{cortes2009, schneider2014, wolff2016}, with the disk being detected along the major axis out to 1.25$\arcsec$ ($\approx$\,124~au) in both $J$ and $H$ band, at which point the signal rapidly drops below the detection limit (c.f. Figure \ref{figSB}). This corresponds to the extent of the CO emission \citep{kastner2010}. The outer ring at 0.8$\arcsec$ is clearly visible in both wavebands. The NE and the South regions of the ring are the brightest in polarized scattered light (with the former being $\approx$\,30$\%$ brighter than the latter). This symmetric enhancement at $30^\circ$ from the disk major axis is most likely entirely due to the maximized polarizing efficiency for a scattering angle of $\approx$\,90$^\circ$, since in disks inclined by $\approx$\,30$^\circ$ this type of scatters roughly occurs at those locations.

The bottom-to-peak contrast between the faint region inside the ring and the ring is on average $\approx$\,40$\%$ in the $H$ band (in agreement with the GPI observations in the same waveband) but somewhat higher in the $J$ band, i.e. $\approx$\,60$\%$. 
Within the faint region, our images seem to reveal a further discontinuity at a distance of $\approx$\,0.6$\arcsec$ (59~au), which is most apparent from the radially scaled image (Figure~\ref{figAll}). Finally, the strongest signal is detected from a compact region of $\approx$\,0.25$\arcsec$ (25~au) in radius. We do not detect any significant inward decrement of signal close to the coronagraph, ruling out the existence of an inner cavity for $\mu$m-sized dust grains larger than $\approx$\,10~au.  

\citet{wolff2016} revealed an azimuthal decrease by 35$\%$ in polarized light at P.A.$=160^\circ-220^\circ$ which is persistent across wavebands and which they ascribe to a shadow cast by a density enhancement at the disk inner edge or by cold spots on the stellar surface. Our images do not reveal such a dramatic dip in the azimuthal distribution, and neither do the HST images, though they are not polarimetric and not taken at the same epoch, but also do not detect variations over two epochs \citep{schneider2014}. We note at this point that an imperfect correction of interstellar or instrumental polarization will lead to butterfly patterns in the $Q_\phi$ and $U_\phi$ images, which overlaid with the disk image can make it appear like there are decrements in the disk. We tried to do a very careful job with our correction for interstellar\,/\,instrumental polarization (see Appendix) and do not see any butterfly patterns in the $U_\phi$ image, indicating that our $Q_\phi$ data is free of such patterns, too. Looking closely at the $Q_\phi$ image presented in \citet{wolff2016}, there seems to be a decrement in the other (northern) direction as well, which would be expected for an overlaid butterfly pattern. Unfortunately, these authors do not show their $U_\phi$ images. We have to also point out, however, that our observations were taken at a different epoch, i.e. almost two years later than the GPI observations.

However, comparing the two bands we have data for, we see a localized dip towards the west (PA$\sim$270$^\circ$) in the $H$ band compared to the $J$ band. At other azimuths, the two bands are comparable. Given the short baseline of only $\approx$\,24 hours between the two observations, this does argue for short-term variations, possibly due to shadowing, in the disk, or variations in the scale height with wavelength of something that can cast a shadow. The signal-to-noise of this feature is low, though, making this detection tentative.

\citet{schneider2014} describe an extended halo outside of the main disk, accounting for $\sim$10$\%$ of the total scattered starlight, which we fail to detect in our images (c.f. Figures \ref{figOverview} and \ref{figAll}). This could be due to the low SNR we can achieve at these large separations and for very faint, optically thin dust, for which HST continues to be very competitive \citep[c.f.][]{schneider2014, olofsson2016}, even though given the fact that we detect a halo out to $>$\,700~au for IM~Lup, this explanation seems unlikely.

\subsubsection{V4046~Sgr}

V4046~Sgr was previously imaged using GPI, the results were presented in \citet{rapson2015a}. Our observations confirm their results - a disk with two rings seen in scattered light - at much higher SNR. We can not, however, confirm their assertion that the scattering efficiency is higher at shorter wavelengths - in fact, the color of the scattering appears to be red, with more light being scattered in the $H$ band (c.f. Table \ref{table:totalContrast}). We can also not confirm the multiple dark lanes seen in their $J$ band image in the inner ring, though we do see a decrement in the $H$ band image at a PA of $\sim$280$^\circ$. This feature likely represents shadowing from the second star of the tight binary and will be discussed in more detail in d'Orazi et al. (in prep.).

We can confirm, through the fitting of rings to the ellipses seen in the image (see section \ref{sec:ringstructures}), that the northwest side is the near side of the disk. We also see that the far side of the disk appears fainter, characteristic for disks at low to medium inclination. This interpretation of near and far side of the disk is also consistent with SMA results from \citet{rosenfeld2012}. These authors estimate the inclination of the disk at 33$^\circ.5^{+0.7}_{-1.4}$. Our estimates are $30.53^\circ\pm0.62^\circ$ for the inner and $32.18^\circ\pm0.51^\circ$ for the outer ring. While this may seem to implicate a trend from smaller to larger inclinations when going from smaller to larger separations and thus a slight warp within the disk, all values agree within 3$\sigma$ w.r.t. their respective error bars. Furthermore, the inclination of the central binary system is determined by the same authors, using unpublished RV constraints, to $33.42^\circ\pm0.01^\circ$. 

The continuum dust emission of the disk can be fit with a characteristic radius of $45^{+5}_{-3}$~au, while the CO disk extends out to almost 400~au \citep{rosenfeld2012, rodriguez2010}. This is consistent with the compact, bright, inner regions with the two rings we detect in the scattered light, while an extended halo is seen out to $\approx$\,250~au and potentially reaching further (see Figure \ref{figSB}), much farther than GPI was able to trace the halo \citep[$\approx$\,45~au,][]{rapson2015a}.

\subsubsection{DoAr~44}

While we do see a bright ring in our DoAr~44 data, with a decrement towards the inside, it is not clear whether this is due to the IRDIS coronagraph or due to an actual decrement (i.e. gap) in the disk that can be resolved. We know from sub-mm ALMA observations that the gas and dust cavities have radii of 19 and 39~au, respectively \citep[0.13$\arcsec$ and $0.27$\arcsec,][]{vandermarel2016}, hinting towards dust filtering and an inner edge of the (gas\,/\,small $\mu$m grains) scattered light rim indeed very close to the coronagraph edge. The ALMA image also shows a ring at significantly larger separations. In scattered light, the disk surface brightness falls off rapidly beyond the bright inner rim, though scattering can be detected above the noise out to $\approx$\,90~au (Figure \ref{figSB}). This again is rather close to the measurements in the sub-mm, where the gas disk can be traced out to $\approx$\,70~au, showing that the disk is overall rather small compared to the disks of, e.g., IM~Lup and RXJ~1615.

Unfortunately, due to the scattered light being so close to the coronagraph, we are not able to determine an inclination for this disk, and are thus not able to confirm the inclination of $\approx$\,20$^\circ$ found in \citet{vandermarel2016}, though we can confirm that the disk looks to be close to face-on from our observations.

Besides that, the disk resembles a scaled-down version of HD~142527 \citep{fukagawa2006, casassus2012, avenhaus2017}, with its bright ring that is broken by two sharp depletions in surface brightness. For HD~142527, these nulls can be explained with an inclined inner disk close to the star, which casts shadows onto the outer disk \citep{marino2015}. {Casassus et al. (submitted) consider a similar scenario for DoAr 44.}

\begin{figure*}
\centering
\includegraphics[width=0.88\textwidth]{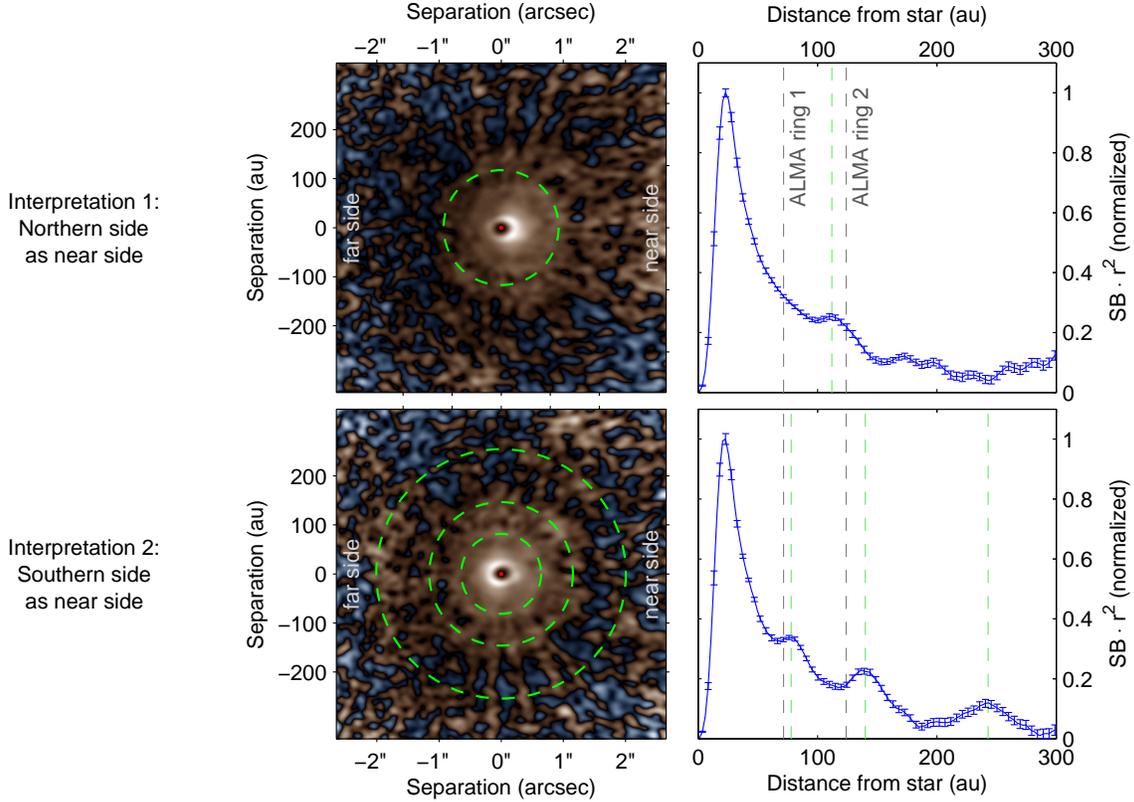}
\vspace{8pt}
\caption{{
The two possible de-projections of AS~209, using $i=35.3^\circ$ and ${\rm PA}=86.0^\circ$, as determined by \citet{fedele2017}, as well as h/r\,(100\,au) = 0.1617 and $\alpha=1.219$ (the average values determined for these parameters in Section \ref{vertStructures}). The first (upper) de-projection assumes the northern side to be the near side, while the second assumes the southern side to be the near side (i.e., ${\rm PA}=86.0^\circ+180.0^\circ$ in our frame of reference). The frames on the left show the de-projected images, displayed in linear stretch after scaling with $r^2$, while the plots on the right show azimuthally averaged surface brightness, also scaled with $r^2$ and normalized to the peak. Rings are detected at around either 120\,au (interpretation 1) or 82\,au, 145\,au, and 255\,au (interpretation 2) and marked with green dashed lines. The ALMA continuum rings seen by \citet{fedele2017} at 75\,au and 130\,au are marked with grey dashed lines. The 1$\sigma$ error bars are calculated from the U$_\phi$ frames, taking into account that the effective beam size changes when de-projecting (see appendix for a detailed description of our error derivations). They do not take into account azimuthal variations in the Q$_\phi$ frame. The dashed lines in the left panel show the position of rings detected in scattered light.}
}

\label{figAS209}
\end{figure*}

\subsubsection{AS~209}

\label{secAS209Results}

The disk of AS~209 at first sight appears similar to RU Lup, in that it is relatively faint and small. Tight ring-like substructures that are easily seen in Figure \ref{figOverview} most likely are not of physical nature, given that they closely resemble the airy ring pattern of the PSF and are furthermore below the resolution of the beam. This is also shown in figure \ref{figAll}, where a re-convolution with a 75~mas beam makes these rings disappear, especially in the $J$ band (faint structures are still seen in the $H$ band).

A possible explanation for these ring-like features at the location of the Airy ring dips is that a bright inner part of the disk, which is known to extend very close to the star \citep{perez2012}, is propagated to larger radii through the PSF. The Airy rings of the PSF can dislocate the flux of a bright inner disk well below the resolution of the telescope (for example, around 0.01\arcsec  from the star) to larger radii, where constructive interference can occur in the Q$_\phi$ band. We tested this with a mockup disk and a perfect Airy ring pattern and were able to produce such rings in Q$_\phi$ at much larger radii than the location of the disk, consistent with our observations of AS~209. No constructive interference was observed in the U$_\phi$ band. This can only occur at high Strehl, where Airy rings are observed. This is consistent with the fact that we do not see such effects in RU~Lup \citep[where the disk extends in very close as well, though indications for a hole on au-scales exist,][]{takami2003} because the AO correction was not good enough to produce a visible Airy ring pattern.

{\citet{andrews2009} find an inclination of $\approx$\,$38^\circ$ along a PA of $\approx$\,$86^\circ$. A more accurate measurement of inclination and position angle ($\approx$\,$35.3^\circ$ and $\approx$\,$86.0^\circ$, respectively) has been provided recently by \citet{fedele2017}. The latter authors also detect rings in the ALMA sub-mm continuum at 72~au and 124~au, along with gaps at 59~au and 98~au, but neither can determine which is the near side of the disk.} 

{Motivated by this, we de-project our disk using the aforementioned parameters \citep{fedele2017}, as well as the average flaring of our disks determined in Section \ref{vertStructures} (h/r\,(100\,au) = 0.1617 and $\alpha=1.219$). We present both possible interpretations, with either the northern or the southern side being the near side, in Figure \ref{figAS209}. Given the low overall SNR of our data, we use the $H$ band and a smoothing kernel of 125~mas. In both cases, we detect surface brightness enhancements, either at 112~au (northern side as near side), or at 78~au, 149~au, and 243~au (southern side as near side). The first interpretation is consistent with the fact that for all our other disks, the near side is the brighter side (see Figure \ref{figure:ringFits}), and the location of the bright spots near the coronagraph on the northern side. The detected scattered-light ring is also just inside the ring in sub-mm detected with ALMA, as is typical \citep[e.g.,][]{garufi2014, bertrang2018}.}

{However, inspecting the image before de-projection (see Figure \ref{figAll}), faint rings are visible in the $H$ band image which are displaced towards the northern side, which argues for interpretation 2, with the southern side being the near side. This is supported by the fact that we detect more rings at higher contrasts. In fact, specifically the inner faint ring visible by eye can still be seen in the first de-projection (interpretation 1, Figure \ref{figAS209}), distorted towards the right side. We also know cases where the far side is the brighter side in polarimetric scattered light, for example HD~100546 \citep{avenhaus2014b, garufi2016} and PDS~70 \citep[][Keppler et al. submitted]{hashimoto2012}. At this point, we cannot determine which interpretation is correct, given that there are valid arguments for either, but in both interpretations the disk displays ring structures. The detection of the rings is relatively robust w.r.t. the flaring angle: The same rings are detected with constant flaring angles between h/r\,=\,0.1 and h/r\,=\,0.3, with no significant changes in their locations. Assuming no flaring (h/r\,=\,0, flat disk), or without performing any de-projection, no rings are detected (see also Figure \ref{figSB}).}

The size of the disk in the mm continuum varies with wavelength, with the disk being larger at shorter wavelengths \citep[$\sim$1$\arcsec$ / 121~au at 0.88~mm,][]{perez2012}. At about this radius, there is a ring-like enhancement in CO emission described by \citet{huang2016}, i.e. at a radius similar to that of the outer ring reported in \citet{fedele2017}. These authors propose CO desorption near the edge of the mm-disk as a possible explanation of this enhancement. While the SNR of our observations decreases rapidly beyond the inner parts of the disk, we trace faint vestiges of scattered light out to $\approx$\,200~au and possibly extending beyond (see Figure \ref{figSB}), meaning that while the disk is unremarkable and relatively faint in scattered light, it is not actually very small compared to the other disks. This is consistent with the fact that CO gas emissions can be traced beyond the diameter of the mm disk, and further supports the notion of the maximum dust grain size decreasing with radius, as $\mu$m-sized grains can be detected at these large radii, while mm-sized grains can not \citep{perez2012, tazzari2016}.

\subsection{Possible companions}

\begin{figure*}
\centering
\includegraphics[width=0.98\textwidth]{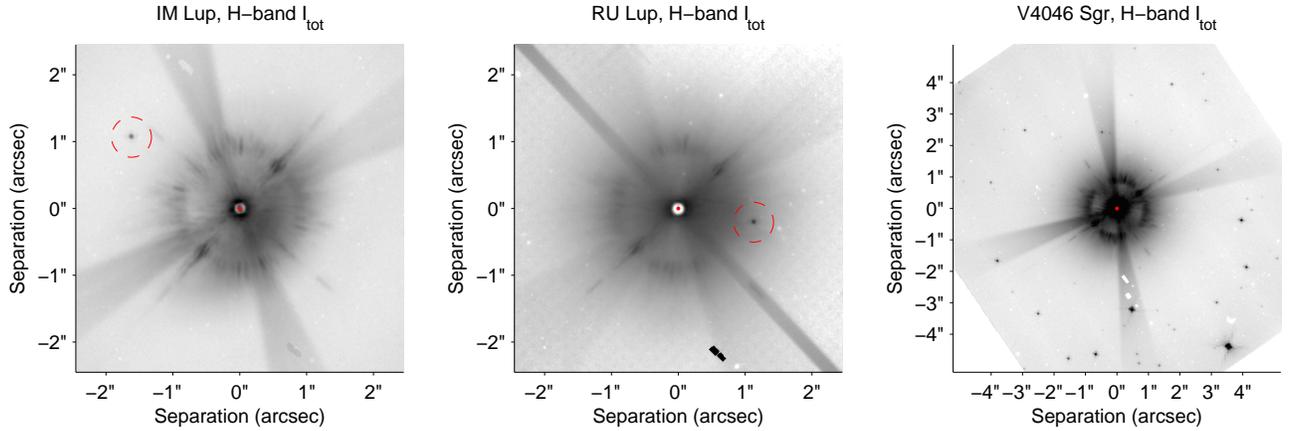}
\vspace{8pt}
\caption{Point source detections in the total intensity images (no PSF subtraction) of our sources. Left and middle: Point sources close to the disks of IM~Lup and RU~Lup (circled in red). Both these point sources were confirmed to not be co-moving using literature or follow-up data. Right: V4046~Sgr showing $>$\,40 point sources in its vicinity, most or all of which are to be expected to be background sources given the position of V4046~Sgr on the sky very close to the galactic plane. The scaling is chosen in all cases to make the point sources most clearly visible. North is up and East is to the left in all three frames.}
\label{figPointSources}
\end{figure*}

While our setup was not optimized for the detection of point sources, we reach a deep background limit ($\approx$\,25 mag in $H$ band and $\approx$\,25.5 mag in $J$ band at 2$\arcsec$ separation, on average) thanks to the good performance of the SPHERE\,/\,IRDIS AO\,/\,coronagraph system and long integration times combined with the fact that our primary targets are relatively faint. Consequently, additional point sources can be seen in the total intensity images of all our datasets, ranging from one or two up to $>$\,40 for the V4046~Sgr dataset. Background sources are to be expected at these magnitudes, specifically in cases like V4046~Sgr, which lies close to the galactic plane. Thus, we expect most, if not all of these point sources to be background objects.

However, two objects seemed particularly interesting: A point source towards the northeast, just outside the disk rim of IM~Lup and an object towards the west at a separation of $\sim$1.1$\arcsec$ of RU~Lup (see Figure \ref{figPointSources}). The point source close to IM~Lup could be shown to be not co-moving using archival NACO data (it does not appear in the archival HST data used to first detect the disk from \citet{pinte2008}, presumably because the point source was behind the disk at that epoch). For RU~Lup, no such archival data was available, but a short exposure one year later showed that the object is most likely a background object.

We did not perform follow-up observations or literature checks for the remaining point sources. However, our data can serve as additional reference points for future studies trying to detect companions to these stars.

\subsection{Correlation of disk and stellar parameters}

Our sample was set up to span a large range in stellar ages, in order to be able to investigate possible evolutionary steps in protoplanetary disks. Our sample also covers a certain range in stellar masses\,/\,spectral types, and comparison to existing studies of Herbig Ae/Be stars \citep[][and many others]{bertrang2018, monnier2017, ginski2016, wagner2015, muto2012} broadens this range considerably.

However, even though all eight disks we observed could be detected, most of them over a large range of radii, and in two different wavelengths, and even though they show a large diversity in structure and physical size, no clear trends with either age or spectral type can be determined. This argues for a scenario in which the formation and evolution of protoplanetary disks and its interplay with planet formation is a complex and chaotic process, in which also other factors such as the formation environment may play important roles, in agreement with theoretical studies of disk evolution \citep{bate2018}. Rather than being examples of a more or less well-defined sequence of evolutionary steps, we might be looking at different evolutionary pathways, for example because some of the disks we investigate are forming (or have already formed) gas giant planets, while others are only forming smaller, rocky worlds \citep{owen2016, cieza2015, williams2011}, even though we do not detect correlations with the sub-mm flux either. This scenario is well in line with the fact that the outcome of planet formation is very diverse, as evidenced by the Kepler results \citep{mullally2015, batalha2013, borucki2010}. {However, we have to keep in mind that the sample was not chosen as an unbiased sample, but based on high (sub-)mm excesses, so it is possible that the older disks represent atypical examples, and that there exist underlying correlations we have yet to discover.}

Comparing our data to existing studies of both TTauri and Herbig stars, there is one peculiarity, though: While spiral features are relatively abundant in Herbig systems \citep{long2017, avenhaus2017, stolker2016, ohta2016, benisty2015, avenhaus2014b, hashimoto2011}, none of our systems shows clear signs of spiral structures. Instead, five out of eight of our systems are clearly dominated by ring structures, with a sixth (AS~209) showing low-SNR ring structures as well. On top of that, DoAr~44 shows a bright inner ring and possibly weaker ring-like structures further out. The only system to not show any ring structures (or any structures whatsoever) is RU~Lup, but this could potentially be due to the fact that the low SNR achieved for this disk does not allow for the detection of any substructures.

\section{Conclusion}
\label{sec:conclusion}

In this work we show the first results of our DARTTS survey and present 8 TTauri disks imaged with SPHERE\,/\,IRDIS in PDI mode at high SNR. All eight disks are clearly detected. The disk show remarkable differences in their total extent and in the amount of substructures they show, with two disks (RU~Lup and AS~209) appearing particularly faint compared to the others. However, there are no significant differences in the 2 filters for each source.

We are able to see the 3-dimensional structure of three of our disks because we detect the lower disk surface (IM~Lup, RXJ~1615 and MY~Lup) and are able to infer the 3-dimensional structure, i.e. the flaring of the $\tau=1$ surface layer, for two more (V4046~Sgr and PDS~66) by means of fitting inclined and elevated rings to their scattered-light images. This way, we can also show that the rings seen in these images are highly consistent with circles (rather than ellipses) that are inclined and displaced horizontally off the disk midplane. We can also show that most TTauri disks seem to follow approximately the same flaring law for their $\tau=1$ surface. {The flaring indices we derive range from $\alpha=1.116\pm0.095$ to $\alpha=1.605\pm0.132$, but it is possible to approximately fit the data for all our sources together with a flaring index of $\alpha=1.219\pm0.026$ and h/r\,(100~au)\,=\,0.1617\,$\pm$\,0.0051.}

This work once again shows the remarkable power of Polarimetric Differential Imaging (PDI), specifically when combined with the power of a high-performance adaptive optics system such as SPHERE. 
High signal-to-noise ratios can be achieved thanks high Strehl ratios in $H$ band even for stars as faint as R\,=\,12 (in medium to good conditions). In this stellar demographic, SPHERE has an advantage compared to instruments such as GPI, which is limited to stars of I~$\lesssim$~9. All our targets are fainter than magnitude 9 in I band. V4046~Sgr is close (I\,=\,9.11) and has been imaged with GPI \citep{rapson2015a}, but our images have significantly higher SNR.

While a full in-depth discussion of all our targets is beyond the scope of this paper, such analyses are already underway and will be published separately (e.g. Casassus et al. (submitted) and d'Orazi et al. (in prep.)). Other targets have already been discussed in detail in the literature \citep[e.g.,][]{deboer2016}.

For a full understanding of our targets, sub-mm observations, specifically high-resolution ALMA data, will be crucial. Efforts to obtain such data are under way under the ALMA sub-part of our survey (DARTTS-A, led by Sebastian Perez). Combined with such observations, the results for the 8 TTauri stars presented in this paper, along with further DARTTS TTauri star observations to be presented in a future paper, will lead to a significant step forward in our understanding of TTauri disks.

\acknowledgments

SPHERE is an instrument designed and built by a consortium consisting of IPAG (Grenoble, France), MPIA (Heidelberg, Germany), LAM (Marseille, France), LESIA (Paris, France), Laboratoire Lagrange (Nice, France),  INAF Ð Osservatorio di Padova (Italy), Observatoire de Gen`eve (Switzerland), ETH Zurich (Switzerland), NOVA (Netherlands), ONERA (France)  and ASTRON (Netherlands), in collaboration with ESO. 
SPHERE also received funding from the European Commission Sixth and Seventh Framework Programmes as part of the Optical Infrared Coordination Network for Astronomy (OPTICON) under grant number RII3-Ct-2004-001566 for FP6 (2004Ð2008), grant number 226604 for FP7 (2009Ð2012) and grant number 312430 for FP7 (2013Ð2016).

HA acknowledges support from the Millennium Science Initiative (Chilean Ministry of Economy) through grant RC130007 and further financial support by FONDECYT, grant 3150643. 
Part of this work has been carried out within the framework of the National Centre for Competence in Research PlanetS supported by the Swiss National Science Foundation. H.A., S.P.Q. and H.M.S. acknowledge the financial support of the SNSF. 
G.H.-M.B. acknowledges financial support from CONICYT through FONDECYT grant 3170657, 
C.C. acknowledges support from project CONICYT PAI/Concurso Nacional Insercion en la Academia, convocatoria 2015, folio 79150049, and financial support from ICM Nucleo Milenio de Formacion Planetaria, NPF. M.B. acknowledges funding from ANR of France under contract number ANR-16-CE31-0013 (Planet Forming Disks). 
C.P. acknowledges funding from the Australian Research Council (ARC) under the Future Fellowship number FT170100040.
S.C. and S.P. acknowledge financial support from CONICYT FONDECYT grant 1171624. Financial support was provided by Millennium Nucleus RC130007 (Chilean
Ministry of Economy). SP acknowledges CONICYT-Gemini grant 32130007.

This research has made use of the SIMBAD database, operated at CDS, Strasbourg, France. This work has furthermore made use of data from the European Space Agency (ESA) mission {\it Gaia} (\url{http://www.cosmos.esa.int/gaia}), processed by the {\it Gaia} Data Processing and Analysis Consortium (DPAC, \url{http://www.cosmos.esa.int/web/gaia/dpac/consortium}). Funding for the DPAC has been provided by national institutions, in particular the institutions participating in the {\it Gaia} Multilateral Agreement.

We thank the staff at VLT for their excellent support during the observations. {We also would like to thank our anonymous referee for comments that helped to improve this paper.}

\facility{VLT:Melipal (SPHERE)}

\begin{appendix}

\section{Data reduction pipeline}

This appendix describes the data reduction pipeline used for the reduction of all data used in this paper, and potentially future papers making use of either SPHERE\,/\,IRDIS or SPHERE/ZIMPOL. It is an updated and improved version of the NACO data reduction pipeline described in \citet{avenhaus2014a}, but we think that the amount of changes to the pipeline implemented since then warrant to describe the entire pipeline in detail again. The goal of the pipeline is to provide the most self-consistent, best-SNR data reduction of the input data possible, and our goal was to be able to reduce all data using (mostly) the same parameters, in order to make the results less parameter-dependent.

\subsection{Data preparation and cosmetics}

There are a total of five types of input frames used for our pipeline: Dark Frames, Flat Frames, Flux Frames, Center Frames, and Science Frames.

As a preparation step, the Dark Frames and Flat Frames are converted to Master Dark Frames and Master Flat Frames using the official ESO SPHERE esorex pipeline recipes. This process also produces BPMs (Bad Pixel Maps), which are later used to identify bad detector pixels. The Master files are then applied to the Flux, Center, and Science Frames in the same fashion. Bad pixels are corrected by filling them with Gaussian-smoothed values from surrounding good pixels. On top of the pixels identified in the BPMs, outliers w.r.t. the local flux ($>$\,\texttt{badPixelSigmaCut}, we use $10\sigma$ here) are also treated as bad pixels in this process. Bad pixels that are more than two pixels away from any good pixel are not corrected, and instead set to \textit{NaN}.

The result are pre-processed and cleaned Flux, Center, and Science frames. These three frames have the following purposes:

\begin{enumerate}
\item{Flux Frames: These frames are taken with the star displaced from the coronagraph. The purpose is to both provide an estimate of the PSF (Point Spread Function) during the observation and to provide a measurement of the flux of the star}
\item{Center Frames: These frames are taken behind the coronagraph, but with a pattern overlaid on the DM (Deformable Mirror) that produces four bright spikes well outside the coronagraph. These spikes can be used to accurately determine the position of the star behind the coronagraph}
\item{Science Frames: These frames contain the actual science data}
\end{enumerate}

\subsection{Determining the position of the star}

The position of the star in both the ordinary and extraordinary beam is determined using the Center Frames mentioned before. In a first step, the \emph{center guess} is roughly determined by smoothing the image with a very large Gaussian kernel and finding the peak. In a second step, the data between a radius of \texttt{centering\_inner\_crop} and \texttt{centering\_outer\_crop} around this \emph{center guess} is extracted (the rest of the data is set to \emph{NaN}), and the median is subtracted to get the background to approximately zero. The image is then Radon-transformed and the peak of this Radon transform is converted back into the location of the star in the image plane. The process is then repeated with the newly determined center used as the \textit{center guess} for a second iteration.

As we take a Center Frame both before and after the observations, we average the position of the star between these two. The difference between the stellar position before and after the Science Frames are taken is usually small, on average 0.23 pixels (2.8 mas) for the $H$-band data and 0.20 pixels (2.5 mas) for the $J$-band data.

\subsection{Reduction of Science Frames}

In a first step, the pre-reduced Science Frames are up-scaled by a factor of \texttt{scaling}, using bi-cubic interpolation. This is done in order to reduce uncontrollable smoothing effects from shifting the images by fractions of pixels further in the data reduction. In this step, the slight difference in pixel scale between the two detector directions (\texttt{IRDIS\_anamorphism}) is also corrected. The data are then centered to the (appropriately scaled) pre-determined position of the star, and fine-centered using cross-correlation between the ordinary and extraordinary beam (this works well because the data are dominated by the unpolarized stellar halo, and the coronagraph produces a sharp edge). After this, the images are aligned to True North using the known True North of the instrument (\texttt{IRDIS\_trueNorth}) and the instrument position angle the data was taken with. At this point, the data are corrected for Dark and Flat Frames, accurately centered, north-aligned and pre-scaled.
In the case of our data for this paper, we checked whether the scaling parameter had any significant effect, and found out that it does not. We thus keep the scaling at 1 for reasons of performance.

\subsubsection{Pre-correction for instrumental polarization}

There are two popular methods that have been used to correct for instrumental polarization in PDI data: Equalizing the flux in the ordinary and extraordinary beam before calculating the Stokes vectors \citep[e.g.][]{avenhaus2014a} and subtracting a polarized halo of the star in order to minimize $U_{\phi}$ after the calculation of the Stokes vectors, as pioneered by the SEEDS team \citep[e.g.,][]{follette2015}. Both these methods assume the star to be intrinsically unpolarized. This is not necessarily justified: Stellar spots can produce intrinsic polarization, and polarized absorption from the disk or a halo can produce polarization. On top of that, there could be interstellar polarization. However, we currently do not know how to a) accurately determine the polarization of the star, and b) how to deal with it if we did. It is worth mentioning several things though: First, the polarization of stars is usually low ($\ll10\%$) compared to the scattering polarization ($15-50\%$) of the disk, so it is likely a second-order effect. Second, interstellar polarization is expected to be low due to the proximity of these stars and furthermore affects both the star and the disk, and as such affects our data in exactly the same way as global instrumental polarization. It would thus just be canceled out by our correction routines.

That being said, we in fact use both the pre-Stokes correction and post-Stokes correction method in our pipeline. The reason is that the post-Stokes correction method can be better fine-tuned and is in general more accurate, but fails in the case of data taken under adverse observing conditions and faint disks (in our case specifically: RU~Lup). We thus use the pre-Stokes correction method (described here) as a means of pre-correction and the post-Stokes correction method (see below) as a means of fine-correcting, combining the strengths of both techniques.

For the pre-correction, the flux in the ordinary and extraordinary beam is measured in an annulus between \texttt{correctInst\_preCorrect\_Rinner} and \texttt{correctInst\_preCorrect\_Router}, and the ratio between the two fluxes is determined. The flux is then equalized by multiplying one of the beams by $sqrt(ratio)$ and the other by $sqrt(1/ratio)$.

\subsubsection{Stokes calculation and stacking}

The Stokes vectors $Q$ and $U$ are calculated in the standard way \citep{tinbergen2005}. The formulas used are:

$$p_{\rm q} = \frac{R_{\rm Q}-1}{R_{\rm Q}+1}\quad;\quad\quad p_{\rm u} = \frac{R_{\rm U}-1}{R_{\rm U}+1}$$

with

$$R_{\rm Q} = \sqrt{\frac{I_{\rm ord}^{0^\circ} / I_{\rm extra}^{0^\circ}}{I_{\rm ord}^{-45^\circ} / I_{\rm extra}^{-45^\circ}}}\quad;\quad\quad R_{\rm U} = \sqrt{\frac{I_{\rm ord}^{-22.5^\circ} / I_{\rm extra}^{-22.5^\circ}}{I_{\rm ord}^{-67.5^\circ} / I_{\rm extra}^{-67.5^\circ}}}$$

Here, the subscripts refer to either the ordinary or extraordinary beam and the superscripts refer to the angular position of the HWP. The Stokes $Q$ and $U$ parameters are then simply calculated as

$$Q = p_{\rm q} * I_{\rm Q}\quad;\quad\quad U = p_{\rm u} * I_{\rm U}$$

where

$$I_{\rm Q} = (I_{\rm ord}^{0^\circ} + I_{\rm extra}^{0^\circ} + I_{\rm ord}^{-45^\circ} + I_{\rm extra}^{-45^\circ})/2\quad;\quad\quad I_{\rm U} = (I_{\rm ord}^{-22.5^\circ} + I_{\rm extra}^{-22.5^\circ} + I_{\rm ord}^{-67.5^\circ} + I_{\rm extra}^{-67.5^\circ})/2$$

\vspace{0.2cm}
are the total intensities in the images used for the calculation of $p_{\rm q}$ and $p_U$. The Stokes vectors for each individual HWP cycle (typically between 4 and 10) are then stacked together using the mean of the individual frames.

\subsubsection{Fine correction for instrumental polarization and Local Stokes Vectors}

The second (post-Stokes) correction for instrumental polarization is integrated with the calculation of the local Stokes vectors $Q_{\phi}$ and $U_{\phi}$, defined as:

$$Q_{\phi}=+Q\,{\rm cos}\,2\theta+U\,{\rm sin}\,2\theta\quad;\quad\quad  U_{\phi}=-Q\,{\rm sin}\,2\theta+U\,{\rm cos}\,2\theta\quad;\quad\quad \theta ={\rm arctan} \frac{x-x_0}{y-y_0}+\gamma$$

The reason these two steps are integrated with each other is that in order to perform the fine correction for instrumental polarization, $U_{\phi}$ needs to be calculated multiple times in an iterative process. Furthermore, during the optimization described below, $\gamma$, which is the correction for a possible mis-alignment of the HWP or otherwise rotated polarization, can also be determined.

The SEEDS team \citep{follette2015} uses a scaled version of the intensity image ($I = I_{\rm Q} + I_{\rm U}$), which is added to the Stokes vectors $Q$ and $U$, in order to minimize the absolute signal in $U_{\phi}$. We expand on this idea, and add constants on top of this, such that the calculation of $Q_{\phi}$ and $U_{\phi}$ becomes:

$$Q_{\phi}=+Q^\star\,{\rm cos}\,2\theta+U^\star\,{\rm sin}\,2\theta\quad;\quad\quad  U_{\phi}=-Q^\star\,{\rm sin}\,2\theta+U^\star\,{\rm cos}\,2\theta\quad;\quad\quad \theta ={\rm arctan} \frac{x-x_0}{y-y_0}+\gamma$$

with

$$Q^\star = Q + c1 \cdot I + c2\quad;\quad\quad U^\star = U + c3 \cdot I + c4$$

This gives a total of 5 variables ($c1$, $c2$, $c3$, $c4$, $\gamma$), over which is optimized in order to minimize the absolute signal in $U_{\phi}$, i.e. $\sum{\lvert U_{\phi}\rvert}$, between an inner and outer radius, \texttt{postStokesCorr\_rInner} and \texttt{postStokesCorr\_rOuter} (listed in Table \ref{table:reductionParameters}). The radii used depend on the geometry of the source, but are kept the same between the two bands for consistency. For the optimization, the MATLAB built-in routine \texttt{fminsearch} is used.

The reason we use a constant on top of the scaled intensity image is that we know that the polarization of the target (due to interstellar or intrinsic polarization) is not necessarily equal to the polarization of the sky background. The constant allows for separate correction of the sky background, and our experiments show that this makes a significant difference particularly for suppressing a butterfly pattern otherwise appearing in  both $Q_{\phi}$ and $U_{\phi}$ at large separations.

{Note that this procedure can in principle remove (astrophysical) signals in the form of butterfly patterns from the U$_\phi$ image. However, the butterfly patterns that can be created by adding constants and multiples of the total intensity frame to the Q and U frames are very limited, and unlikely to match astrophysical signals well. Indeed, the butterfly-like signal in the MY~Lup U$_\phi$ frame (see Figure \ref{fig:SelfCancelCorrection}), though not astrophysical in nature (see next paragraph) is not removed.}

To convert the images to a physical scale, we use the \texttt{IRDIS\_pixelscale} of 12.258\,mas/pixel and the distances to our targets acquired from the literature.

\begin{deluxetable}{lc}
\centering
\tablewidth{0pt}
\tablecaption{Reduction parameters
\label{table:reductionParameters}}           

\tablehead{
\colhead{Parameter} & \colhead{Value}
}
\startdata
\texttt{badPixelSigmaCut} & $10\sigma$ \\
\texttt{centering\_inner\_crop} & 400\,mas (J band) / 500\,mas (H band) \\
\texttt{centering\_outer\_crop} & 500\,mas (J band) / 670\,mas (H band) \\
\texttt{scaling} & 1 \\
\texttt{IRDIS\_anamorphism} & [1.006 1] \\
\texttt{IRDIS\_trueNorth} & -1.775$^{\circ}$ (East of North) \\
\texttt{correctInst\_preCorrect\_Rinner} & 0\,mas \\
\texttt{correctInst\_preCorrect\_Router} & 1000\,mas \\
\texttt{IRDIS\_pixelscale} & 12.258\,mas/pixel \\
\enddata
\tablenotetext{}{Parameters used in data reduction for this paper. The negative angle for True North means that in order to correct for True North, the images have to be rotated clockwise by 1.775$^\circ$.}
\end{deluxetable}

\begin{deluxetable}{lcc}
\centering
\tablewidth{0pt}
\tablecaption{Instrumental polarization correction radii
\label{table:postPolCorrectionParameters}}           

\tablehead{
\colhead{Target} & \colhead{\texttt{postStokesCorr\_rInner}} & \colhead{\texttt{postStokesCorr\_rOuter}}
}
\startdata
IM~Lup		& 0.5 & 3.5 \\
RXJ~1615	& 0.0 & 3.5 \\
RU~Lup		& 0.0 & 3.5 \\
MY~Lup		& 1.0 & 3.5 \\
PDS~66		& 0.5 & 3.5 \\
V4046~Sgr	& 0.0 & 3.5 \\
DoAr~44		& 0.5 & 3.5 \\
AS~209		& 0.5 & 3.5 \\
\enddata
\tablenotetext{}{Inner and outer correction radii for the second part of the instrumental polarization correction (in arcseconds).}
\end{deluxetable}

\subsection{Correction for PDI self-cancellation and Q$_\phi$\,/\,U$_\phi$ cross-talk}

\begin{figure}
\centering
\includegraphics[width=0.98\textwidth]{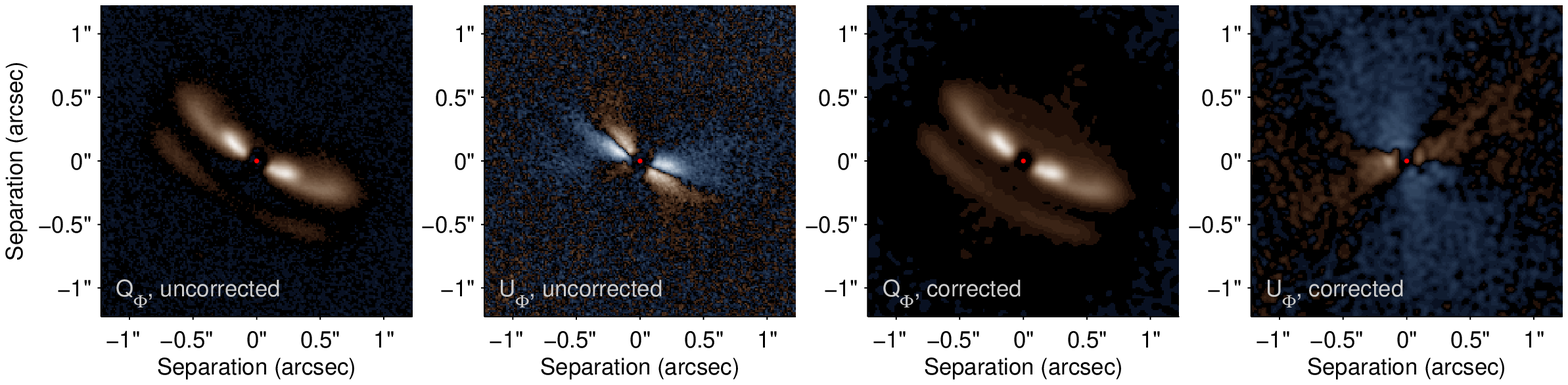}
\vspace{8pt}
\caption{An example of the effects of self-cancellation and Q$_\phi$/U$_\phi$ cross-talk correction (from left to right: $Q_\phi$, $U_\phi$, $Q_{\rm \phi,\,corrected}$, $U_{\rm \phi,\,corrected}$). While the correction on $Q_\phi$ mostly has the effect of making the disk brighter, especially in the inner regions, the correction on $U_\phi$ removes the {strong} butterfly-like pattern, which could be misinterpreted as a residual signal from multiple scattering, similar to what is expected theoretically \citep{canovas2015}, {though with the opposite sign, which could potentially help to distinguish these effects}. The $U_\phi$ images have been scaled up by a factor of 5 to make the fainter signal w.r.t. $Q_\phi$ visible. Blue hues correspond to negative, brown hues to positive values. {While the total flux in $U_\phi$ remains largely unchanged with the mean being around zero, the standard deviation is significantly reduced, by a factor of 2.14.} North is up and East is to the left.}
\label{fig:SelfCancelCorrection}
\end{figure}

At this point, the corrected data is still affected by two effects that we know of, which are inherent to PDI and can lead to misinterpretation of the data: PDI self-cancellation and Q$_\phi$\,/\,U$_\phi$ cross-talk.

PDI self-cancellation has been discussed before and stems from the fact that close to the position of the star, the positive and negative signals from the butterfly patterns in both the Stokes Q and the Stokes U vector cancel each other out due to being smeared out by the PSF of the telescope. A description of this effect, together with an example, can be found in \citet{avenhaus2014b}.

Q$_\phi$/U$_\phi$ cross-talk has, to our knowledge, not been discussed before. The decomposition into Q$_\phi$ and U$_\phi$ relies on knowledge of the position angle in the field, but the flux from one point in the sky is distributed over an area due to the PSF. This leads to the usage of an incorrect position angle for the decomposition, which in turn leads to incorrect results. Unfortunately, the patterns produced by this effect can, especially for highly inclined disks, resemble the patterns theoretically expected from multiple scattering effects in such disks, which makes them prone to misunderstanding. The effect clearly shows up in our data on V4046~Sgr, as seen in Figure \ref{fig:SelfCancelCorrection}.

While an exhaustive discussion of possible corrections for these effects will be presented in a separate paper (Avenhaus et al. in prep.), we want to briefly describe how we deal with the effect here. Both effects stem from the fact that the resolution in our images is finite and limited by the telescope PSF. If we could image Q and U at infinite resolution, neither of the effects would occur.

Thus, we use the following method to correct our data: First, we deconvolve the corrected Q and U data using Wiener deconvolution, using the PSF obtained from the flux frames. These data then have very high resolution and very low SNR, because deconvolution strongly increases the noise. From these Q and U frames, we calculate Q$_\phi$ and U$_\phi$, which both are also unusably noisy. We then re-convolve these images, which brings the resolution back to the level we had before while also bringing the noise back to a similar level. Assuming we use the right PSF, this process brings us very close to the actual signal (in theory, in the absence of noise and without a coronagraph, the reconstruction is perfect). Another benefit is that we do not have to use the same PSF to re-convolve, but can choose another PSF (for example one with finite support, or a Gaussian of known FWHM).
In addition to the original PSF, we use Gaussians with FWHMs of 50, 75, 100, and 125\,mas for this purpose, allowing us to compare the disks at similar resolution.

As far as we can tell (also by applying this technique to simulated data), this method works remarkably well, but depends on the quality of the PSF. As can be seen in Figure \ref{fig:SelfCancelCorrection}, this affects both the $Q_\phi$ and $U_\phi$ signals. While the $Q_\phi$ signal is mostly getting suppressed by the self-cancellation effect with the Q$_\phi$/U$_\phi$ cross-talk having no perceptible impact, the $U_\phi$ clearly is affected by Q$_\phi$/U$_\phi$ cross-talk (prone to possible mis-interpretation in terms of multiple scattering). It is thus of vital importance to understand this effect, and correct for it.

\subsection{Error estimation}

\begin{figure*}
\centering
\includegraphics[width=0.99\textwidth]{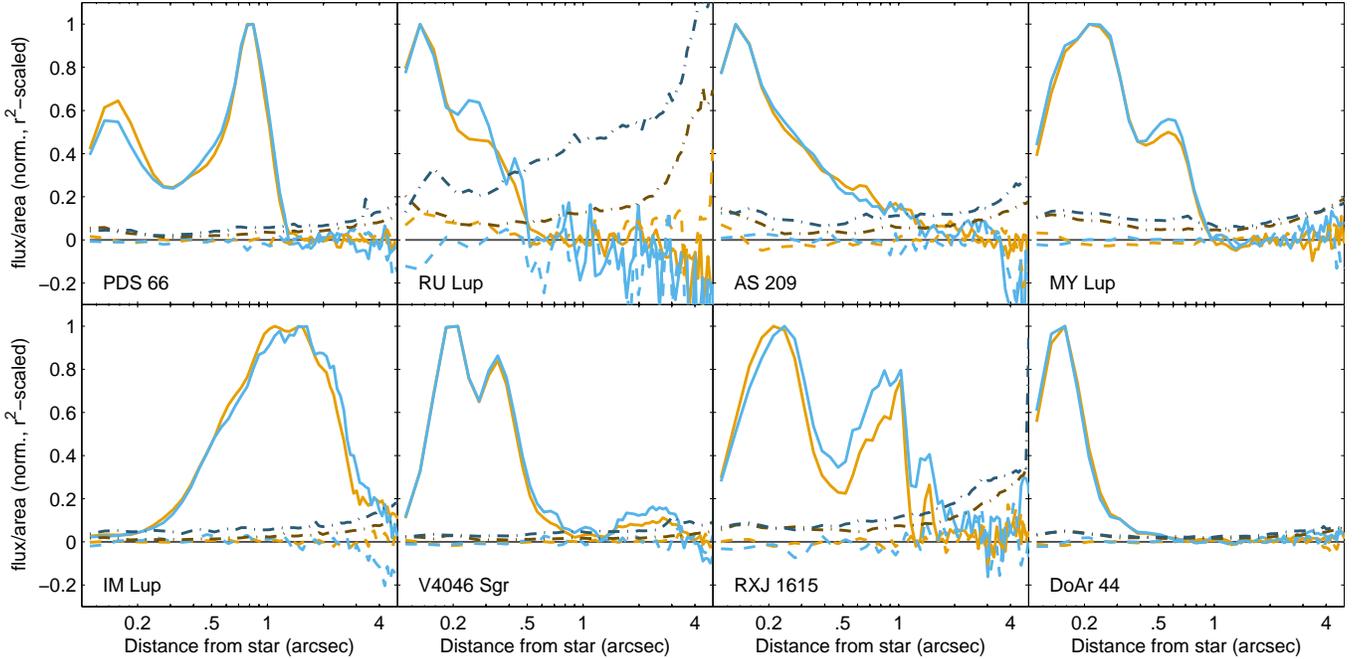}
\vspace{8pt}
\caption{{Individual, normalized Q$_\phi$ and U$_\phi$ surface brightness plots vs. distance for all our targets. Orange lines present $H$ band, blue lines $J$ band data. Solid lines represent Q$_\phi$, dashed lines represent U$_\phi$. Darker, dash-dotted lines represent 3-$\sigma$ detection limits. The width of the annuli used for averaging increases with radius proportional to r$^{1/2}$ (similar to Figure \ref{figSB}). Note the logarithmic stretch of the x-axis. The data have been scaled with r$^2$ in order to improve readability.}}
\label{figSBAll}
\end{figure*}

\begin{figure}
\centering
\includegraphics[width=0.45\textwidth]{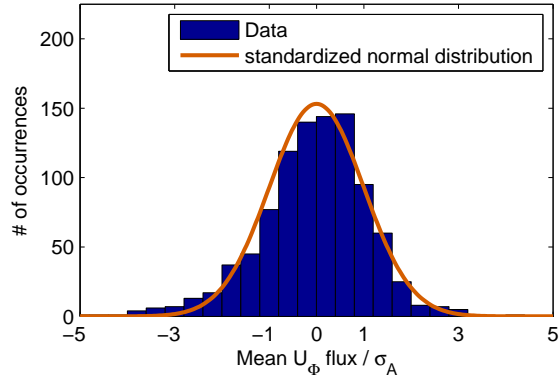}
\vspace{8pt}
\caption{{Mean $U_\phi$ flux across all annuli used to calculate the surface brightnesses in Figure \ref{figSBAll}, divided by the respective error estimate $\sigma_{\rm A}$. The data follows a standardized normal distribution reasonably well, with some additional outliers.}}
\label{figUphiErrors}
\end{figure}

{Since both the $Q_\phi$ and the $U_\phi$ frame are affected by noise (readout noise, speckle noise, systematics) in a similar way, but $U_\phi$ is largely devoid of signal, we use the variance in the $U_\phi$ frame in order to estimate the noise. If there is any astrophysical signal in the $U_\phi$ frame, this method will be conservative, because it over- rather than underestimates the noise in this case.}

{Throughout this paper, we estimate errors for image areas (such as annuli around the star), rather than for point sources. This means that we can take advantage of the fact that errors will tend to average out over larger image areas. Our standard deviation for the mean flux in an image area $A$ thus becomes:}

$$\sigma_{\rm A} = \sqrt{\frac{var(U_{\phi, \rm A})}{N}}$$

\noindent {where $var(U_{\phi, \rm A})$ is the variance in the $U_\phi$ frame over the respective area, and $N$ is the number of resolution elements in the area. This method ensures that our error estimates are independent of the kernel we use for re-convolution (see above), as wider kernels will lead to stronger smoothing, but also to fewer resolution elements.}

{Given that we correct $U_\phi$ to be zero on average, we can then expect the distribution of mean fluxes in $U_\phi$ across many areas, divided by their respective $\sigma_{\rm A}$, to approximately follow a standardized normal distribution. We use this as a sanity check. We show that this is approximately correct in Figures \ref{figSBAll} and \ref{figUphiErrors}. The mean signal of $U_\phi$ across a total of 960 areas (annuli) surpasses 3$\sigma$ only three times. The distribution does follow the standardized normal distribution reasonably well, with only few additional outliers, even though we do not correct $U_\phi$ over the entire separation range (0.1$\arcsec$-5.2$\arcsec$), but only between the inner and outer radii specified in Table \ref{table:postPolCorrectionParameters}. However, the Shapiro-Wilk-test clearly shows that this is not a Gaussian distribution (p-value $\approx\,1.5\cdot10^{-9}$). The outliers are mostly from the very inner and outer regions: Restricting our analysis to separations between 0.5$\arcsec$ and 3.5$\arcsec$, the distribution of $\overline{U_{\rm\phi,A}}$/$\sigma_{\rm A}$ is indistinguishable from a Gaussian distribution (576 annuli, p-value $\approx\,0.2$).}

\begin{figure}
\centering
\includegraphics[width=0.90\textwidth]{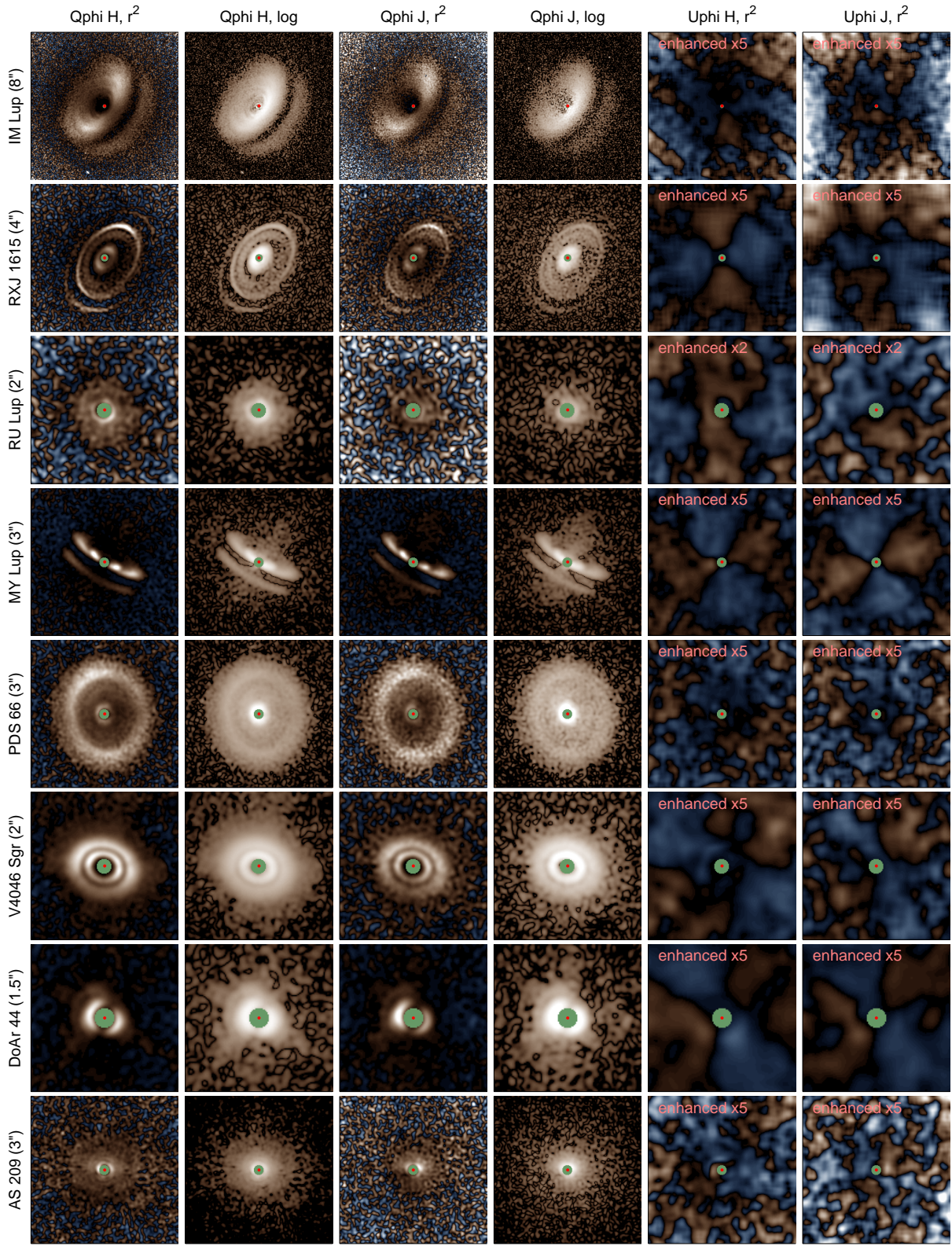}
\vspace{8pt}
\caption{All our observations, corrected for self-cancellation as described in the appendix and re-convolved with a 75\,mas FWHM Gaussian. The horizontal/vertical FOV is given with the name of the disk. Blue hues correspond to negative, brown hues to positive values. North is up and East is to the left in all frames.}
\label{figAll}
\end{figure}

\vspace{1cm}

\end{appendix}



\bibliographystyle{aasjournal.bst}
\bibliography{DARTTS_I.bib}



\end{document}